\documentclass[final,12pt,authoryear]{elsarticle}
\makeatletter
\def\ps@pprintTitle{%
     \let\@oddhead\@empty
     \let\@evenhead\@empty
     \def\@oddfoot{\hbox to \textwidth{\ifnopreprintline\relax\else\@myfooterfont\ifx\@elsarticlemyfooteralign\@elsarticlemyfooteraligncenter\hfil\@elsarticlemyfooter\hfil\else\ifx\@elsarticlemyfooteralign\@elsarticlemyfooteralignleft\@elsarticlemyfooter\hfill{}\else\ifx\@elsarticlemyfooteralign\@elsarticlemyfooteralignright{}\hfill\@elsarticlemyfooter\else\ifx\@journal\@empty Elsevier\else\@journal\fi\hfill\@date\fi\fi\fi\fi}}%
     \let\@evenfoot\@oddfoot}
\makeatother

\usepackage{amssymb}
\usepackage{amsmath}
\usepackage{wasysym}
\usepackage{mathrsfs}
\usepackage{booktabs}
\usepackage{soul}
\usepackage{xcolor} % macro color names
\usepackage[ruled,linesnumbered,algonl]{algorithm2e}
\usepackage{tikz}
\usetikzlibrary{tikzmark,decorations.pathreplacing, calc}

\SetKwInOut{Definitions}{Definitions}
\SetKwInOut{Parameters}{Parameters}

\usepackage{lineno}
\journal{}

\begin{document}

\begin{frontmatter}


\title{Incremental Seeded EM Algorithm for Clusterwise Linear Regression}

\author[Waikato]{Ye Chow Kuang}
\author[Waikato]{Melanie Ooi}

\affiliation[Waikato]{organization={University of Waikato},
            addressline={Private Bag, 3105}, 
            city={Hamilton},
            postcode={3240}, 
            state={Waikato},
            country={New Zealand}}

\begin{abstract}
This paper proposes Incremental Seeded Expectation Maximization, an algorithm that improves upon the traditional Expectation Maximization computational flow for clusterwise or finite mixture linear regression tasks. The proposed method shows significantly better performance, particularly in scenarios involving high-dimensional input, noisy data, or a large number of clusters. Alongside the new algorithm, this paper introduces the concepts of \textit{Resolvability} and \textit{X-predictability}, which enable more rigorous discussions of clusterwise regression problems. The resolvability index is quantified using parameters derived from the model, and results demonstrate its strong connection to model quality without requiring knowledge of the ground truth. This makes the \textit{Resolvability} especially useful for assessing the quality of clusterwise regression models, and by extension, the conclusions drawn from them.

\end{abstract}

%\includegraphics{xxx}

% \begin{highlights}
% \item EM algorithm optimized for clustered-regression type problems; greatly improved the probability of finding the global optimum even for traditionally challenging classes of clusterwise linear regression problems. 
% \item Proposed ``cluster resolvability'',  a useful conceptual tool and a quantitative measure of clusterwise linear regression model quality. 
% \item  Proposed ``X-predictability'',  a useful conceptual tool and a quantitative measure that enables a better framework to report clusterwise linear regression model predictions.
% \end{highlights}


\end{frontmatter}

% Disable hyphenation
\hyphenpenalty      = 1000
\exhyphenpenalty    = 1000
% Optional: Adjust text stretching
\sloppy
% Enable line number
%\linenumbers

\section{Introduction}                  \label{sec: introduction}
Regression models are widely used in research across many fields of study relating explain dependent (or response) variable $y$ to independent (or explanatory) variables $X$. Of all the regression methodologies, the most well-studied and widely applied regression models is the multiple linear regression of the form $y=X\beta$ where $\beta$ is a regression vector encoding the contribution of each factor to the response. The linear regression model forms the fundamental module on which many more sophisticated regression algorithms are being built. \textit{Clusterwise linear regression} (CLR) is one of the generalizations to regression that was proposed independently multiple times in the literature. The needs of clusterwise regression analysis naturally arises when data exhibits heterogeneity and was found to have been applied in to all endeavors of human investigations covering social studies, economics, biology, meteorology, engineering and medicine, see \citep{RN28} and \citep{Kuang} and references there in. Data is said to be heterogeneous when unknown subgroups exist within the dataset and each subgroup behave according to distinct local functional model between the response and explanatory variables \citep{RN42}. The goal of clusterwise regression is to discover the cluster structure and regression function simultaneously within the same computational loop; specifically CLR further restricts the search to linear regression functions.  This paper focuses on algorithmic improvement of underlying \textit{Expectation-Maximization} (EM) algorithm as applied on CLR problems and introduces new concepts that improves clarity in the discussion of CLR problems. The proposed improvements are, in principle, applicable to other finite mixture or clustered models related to CLR problems. However, to keep the presentation succinct, the discussions to these generalizations will be kept to minimal.  \par

\subsection{Problem Formulation}                  \label{sec: formulation}

Given a dataset as $D=(X,y)$ where $X \in \mathbb{R}^{N \times p}$ and $y\in \mathbb{R}^{N \times 1}$, $p$ being the dimension of the input data and $N$ the number of independent observations. CLR algorithm will assign each pair $(X_n,y_n)$ of observation to the best subpopulation (or cluster) $D_k$, such that $D = \bigcup_{k=1}^{K}D_k$; and each of the cluster $D_k$ has an optimal regression vector that is distinct. For each observation $(X_n,y_n)$ in $D_k$, one has the linear relationship $y_n = X_n \check{\beta}_k + \beta_{k0} + \varepsilon_k$ where $\varepsilon_k$ is a stochastic variable representing the modeling noise, $\check{\beta}_k$ the cluster specific regression vector and $\beta_{k0}$ the cluster specific offset. It is important to retain $\beta_{k0}$ as a free parameter in a  CLR model because, unlike regression of homogeneous data, these offsets for each cluster cannot be removed by data centering. $\beta_{0k}$ should only be removed if there is a strong, context-specific reason to do so; otherwise, omitting them will substantially diminish the modeling capacity of the CLR models. The augmented variable $[1,X]$ of $X$  will be denoted $\tilde{X}$ in the rest of the paper to enable more succinct expression, as shown in equation \ref{equ1}. Similarly, a regression vector $\beta$ is sometimes broken up into $ \beta_0$ and $\check{\beta}$ where $\check{\beta}$ denotes the part of the vector without the offset component.  

\begin{equation} \label{equ1}
    y = \tilde{X} \beta_k + \varepsilon_k
\end{equation}

Equation \ref{equ1} is incomplete formulation of CLR problem until the cluster membership of each observation is assigned. This can be achieved through a membership weight matrix $w\in \mathbb{R}^{N \times K}$. If the \textit{n}-th observation $(X_n,y_n)$ belongs to $D_k$, then $w_{n,k} = 1$, while other entries of \textit{n}-th row are zero $w_{n,j \neq k} = 0$. With this, equation \ref{equ2} re-states CLR in a more compact form. The norm in equation \ref{equ2} is intentionally non-specific so that the formulation is compatible with various likelihood models that are optimized for different distributions of $\varepsilon_k$.  $\varepsilon_k$ will be assumed to be normally distributed with zero mean in this paper. 

\begin{equation} \label{equ2}
    \arg\min_{\beta_k , w_{n,k}}  \sum_{n=1}^{N}{ \sum_{k=1}^{K}{ w_{n,k} \Vert y_n - \tilde{X} \beta_k \Vert }}
\end{equation}

One can easily see from equation \ref{equ2} that $w_{n,k}$ need not be discrete values \{0,1\} as long as each row $w_{n,\cdot}$ satisfies the constraint $\sum_{k=1}^{K} w_{n,k} = 1 \quad \forall n$. This formulation can be found in works by \citep{RN34,RN33} is more commonly called the \textit{Finite Mixture Linear Regression} (FMLR). FMLR is, in principle, more flexible than CLR for allowing fuzzy cluster membership assignment. All algorithms developed for FMLR, with small modifications, are applicable to CLR and vice versa. A large-scale randomized test involving many classes of problem performed by \citep{Kuang} suggests that no significance difference in median convergence rate and the quality of the solutions between CLR and FMLR. It is the context of application that determine if it is necessary to enforce crisp class membership. In the following discussions, fuzzy cluster membership is adopted and no distinction will be made between CLR and FMLR. However, considering the word "cluster" in the name offers a more intuitive geometric grasp of the problem, the term ``CLR'' will be used to represent both models. It is worth noting here that the CLR formulation described here can be viewed as an algorithm find model of data generated by mixture of different processes; it is also equally valid to consider it an algorithm to find a piece-wise linear approximation to nonlinear functions. \citep{RN31} proposed K-plane regression through CLR with a slightly modified objective function. More on this in Section \ref{sec: resolvability}\par

\subsection{Algorithms}                 \label{sec: algorithms}

The backbone of algorithms to solve CLR problem is the EM algorithm initially proposed by \citep{RN2}. The original algorithm is sometimes called the Späth's algorithm with various minor modifications introduced later to improve its performance. A generic EM algorithm to solve CLR problems is shown in Algorithm \ref{alg: EM}. Algorithm \ref{alg: EM} is sufficiently general to represent the Späth algorithm and most later added innovations. 

\begin{algorithm}[t]
\caption{Generic EM computational flow for CLR problems}\label{alg: EM}
\KwIn{$\tilde{X}$, $y$} 
\KwOut{$w$, $\hat{\beta}$, $\hat{\sigma}$} 
\Parameters{$\zeta$}
\Definitions{$\hat{\beta}$ = [$\hat{\beta_1},...,\hat{\beta_K}]$; $\hat{\sigma}$ = [$\hat{\sigma_1},...,\hat{\sigma_K}$]; $k \in \{1..K\}$ }

Initial guess $\hat{w}$, $\hat{\beta_k}$, $\hat{\sigma_k}$ \\
\While{($\neg$ \textnormal{Converged}) $\wedge$ ($i <$ \textnormal{MaxLoop})}{
    % \tcp*[f]{E-steps} \\
    \textnormal{Evaluate fitting error} $\varepsilon = y-\tilde{X}\hat{\beta}$ \\
    \textnormal{Re-estimate scale parameter} $\hat{\sigma_k}$ \\
    $z^2 = \varepsilon^2/\hat{\sigma}^2$ \\
    \textnormal{Re-weight data} $\hat{w} \leftarrow \phi(z)$ \\
    
    % \tcp*[f]{M-steps} \\
    \textnormal{Estimate the new} $\hat{\hat{\beta_k}}$ \textnormal{using the new} $\hat{w}$ \\
    \textnormal{Momentum update} $\hat{\beta_k} \leftarrow \zeta\hat{\beta_k} + (1-\zeta) \hat{\hat{\beta_k}}$ \\
    \textnormal{Keep track of the best models} \\
    $i \leftarrow i+1$
}
\end{algorithm}

The algorithm starts with initialization step 1.1. When prior solutions or knowledge are absence, random cluster assignment is often used in this step. The literature widely acknowledges that the algorithm may fail to find an optimal solution, often getting trapped in local minima. The Multi-start Späth method, where the EM algorithm is run multiple times with random initializations, is frequently employed as an ad-hoc response to the local minimum issue. \citep{RN7} focuses on finding good initial solutions, demonstrated that the EM algorithm exhibits geometric convergence once the selected initial solution falls within the basin of attraction. This initialization step can be highly complex in some instances. For example, non-EM solutions to CLR problem such as \citep{RN12} and \citep{RN13} proposed to solve CLR problem as tensor equations. The numerical instability caused by high-order moment estimation and approximations procedures made to reduce computational complexity often lead to large error in the solutions. To ensure reliable convergence, the solutions produced by tensor factorization method were feed to the EM algorithm to guarantee the quality of the final solutions. \citep{RN16} proposes the initial values through a custom ``incremental initialization'' procedures which applies some heuristic to guess the initial values and discovers the cluster one-at-a-time.  \par

With the cluster membership initialized, step 1.3 evaluate the regression error while step 1.4 re-estimate the scale-parameter withing each cluster. step 1.5 computes the normalized error $z$ for each data point. This $z$ is then substituted into a likelihood function $\phi$ of choice to reassign/revise the cluster membership in step 1.6. Armed with the new weight $w$ which measures the cluster affinity, step 1.7 re-estimate regression vector $\beta_k$ for each cluster. $\phi(z)$ is traditionally chosen to be a normal density function but other functions can be used. For example \citep{RN11} uses Student-t density function to improve robustness with respect to outliers. The drawback of using Student-t density function is that the hyper-parameter of $\phi$ needs to be re-estimated every cycle as the cluster membership $w$ changes, making its use computationally expensive. \citep{RN39} imposes additional data-driven constraints during the estimation of $\sigma_k$ to avoid numerical singularity in the evaluation of $\phi$. \citep{RN4,RN9,RN10,RN11} reported improved stability when robust estimators are being used in step 1.4 and step 1.7. Maximum likelihood estimation, least-square regression, sparse regression \citep{RN5} and partial leastsquare regression \citep{RN3} all can be applied in step 1.7 to estimate $\hat{\beta_k}$. \par

Finally, Step 1.8 updates the estimated regression vectors $\hat{\beta_k}$. The steepest descent approach described in the literature, where $\hat{\hat{\beta_k}}$ directly replaces $\hat{\beta_k}$, can be recovered by setting $\zeta = 0$. By introducing a momentum update, the algorithm gains additional flexibility to better adapt to the requirements of different problems. It is important to note that the EM algorithm applied to the CLR problem does not guarantee a monotonically decreasing regression error. In complex CLR problems, it is not uncommon for better solutions to be lost during the iteration process, potentially resulting in suboptimal outcomes. To address this issue, Step 1.9 has been introduced to ensure that the algorithm reports the best-performing model found across all iterations. \par 

\subsubsection{Other Algorithms}  \label{sec: other algorithm}

Other types of CLR algorithms that do not rely on EM algorithm directly can be grouped into two categories: tensor formulation and reformulation of CLR to suit generic optimization. Tensor formulation presented by \citep{RN12} and \citep{RN13} are very refreshing attempts to solve CLR. Unfortunately the tensor formulation is plagued with high computational complexity and numerical instability sometimes leading convergence to the wrong solutions. The latest result following this line of investigation ended as initialization method because EM algorithm is needed at the end to verify the solution. \citep{Kuang} recently demonstrated that the tensor formulation induced initial solutions do not offer statistically significant accuracy advantage compared to random initialization. Worse still, the solutions provided tensor formulation is only useful when the clusters are  heavily overlapping. The tensor methods fare worst when the clusters are well-separated likely due to lost of numerical accuracy in the high-order moment computations. \par

Another research direction is to reformulate the CLR problem into a form that is compatible with off-the-shelf global optimization algorithms. \citep{RN18} and references therein represents the effort in this direction. These reformulation always accompany by dramatic increase in the number of free parameters. Even though the global optimization algorithm is more sophisticated than EM algorithm, the complexity of the problem to be solved has increased simultaneously. The high degrees of freedom created more local minima in a higher dimensional search space. Solving CLR directly through metaheuristic optimization technique such  as simulated annealing \citep{RN6} and genetic algorithms \citep{RN46} have also been attempted. \citep{RN15} proposed an algorithm that solves CLR problems by separating clusters sequentially. Their method uses gradient descent instead of regression, and uses k-GMM to resolve cluster membership instead of the likelihood function $\phi(z)$. However, the core k-GMM is fundamentally a re-statement of the CLR problem assuming some clusters have been correctly identified. One major weakness of the algorithm is its reliant on high-order moments, leading the algorithm to have low data efficiency and the performance scale badly with the number of cluster $K$. \par

Finally, there are \textit{subspace clustering} algorithms developed by computer vision community to do image segmentation. The subspace clustering problem, in fundamental terms, a variant of CLR formulation. However, most of these algorithms are are designed to exploit the characteristics of computer vision problems, such as the spatial relationship between data points, extremely high-dimensional data and availability of low-dimensional representations of the original data. Hence, these methods are not included in the review of general CLR algorithms. The exception to this rules are the generalized principal component analysis \citep{RN51} and K-flat algorithms \citep{RN50, RN43,RN20}. The generalized principal component analysis is a slightly different approach to solve the tensor formulation mentioned above with very similar drawback on computational complexity. The K-flat algorithms are variants of EM-algorithm, these algorithms will be revisited in Section \ref{sec: new beta}.

\subsubsection{Challenges}  \label{sec: challenge}

Large-scale controlled randomize evaluation of various EM-based CLR algorithms is performed by \citep{Kuang} to understand the strengths and weaknesses of various proposals to solve CLR problem found in the literature. The study reveals several serious shortcomings and interesting observations about the use of EM-algorithm in CLR.  This paper proposes an improved EM algorithm to address these shortcomings and propose a framework to discuss CLR problems that is less prone to ambiguity. The main findings from \citep{Kuang} are summarized below:

\begin{enumerate}
    
    \item Some performance improvements demonstrated in the past using small problem-set or hand-pick examples turn out to be statistically insignificant when tested on a randomly generated large problem-set. The randomize testing methodology of \citep{Kuang}, summarized in Section \ref{sec: methods}, will be adopted in this paper. 
    
    \item Problem with small inter-clusters distance (between cluster centroids) are consistently more difficult to solve than problems with large inter-cluster distance. Problems with large inter-clusters distance can be solved easily by performing clustering first and then followed by regression. The CLR problem becomes more challenging the smaller inter-cluster separation. CLR problems as a whole, and those with small inter-cluster distances specifically, are sensitive to data corruption. To be succinct, only problems with zero inter-clusters distances will be used in this paper because it represents the most challenging subset of CLR problems. Section \ref{sec: resolvability} and \ref{sec: making predictions} introduce the concept of \textit{resolvability} and \textit{X-predictability} of a CLR problem to facilitate clear discussion about different ``grades'' of CLR problems. 
    
    \item \citep{RN7} has also shown that the CLR problems are NP-hard. The CLR problems are difficult precisely because there is no known fast algorithm to solve the problem; but the efforts to improve the algorithm by approximating a good initial solutions are brute-force attempts to do exactly that. It was found that complex procedures to find a ``good'' initial solution perform worse than K-mean initialization in many cases. K-mean initialization is recommended for its speed, simplicity and good performance, especially when inter-cluster distances are large. K-mean initialization performs on-par with random initialization when the inter-clusters distances are small, no initialization method statistically outperform random initialization in this case. 
    
    \item The intuitions obtained from $K=2$ does not translate easily to cases with $K>2$ due to stronger interactions between clusters during optimization. E.g. robust estimators that confers small (but statistically significant) robustness improvement when $K=2$, produces negligible advantage when $K>2$. It is worth noting here that EM$_{is}$ proposed in Section \ref{sec: isem} makes significant more improvement, that the use of robust estimators is unnecessary even in the case of $K=2$. 

    \item The traditional CLR algorithms can only solve very limited set of simple problems. The convergence rate deteriorate rapidly with increasing $K$ and $p$. The median performance can be bad for problem as modest as $K=3$, $p=10$ (see Section \ref{sec: results}). This is caused by EM algorithm being trapped by some local minima. Once trapped, unable to escape by making use of the available information in the dataset. The performance decreases regardless of the type of initialization method, type of robust estimator or type of likelihood function $\phi(z)$ used. 
    
    \item Using multiple restart to address local minimum problem is computationally inefficient. The computational time increases exponentially while the rate of correct convergence improves very slowly. The improvement quickly becomes negligible when the probability of being trapped in local minima increases with $K$ and $p$. 
    
\end{enumerate}

The proposed EM$_{is}$ algorithm in Section \ref{sec: isem} made an attempt to address the last two issues and demonstrate empirically that the proposals remove significant performance bottleneck. 

\subsection{Related Works}                  \label{sec: related works}
CLR is not the only possible generalization of the conventional linear regression methodology to heterogeneous data. Many of these extensions shared the same EM algorithm as their computational backbone, therefore the proposed CLR specific version of EM is likely to be compatible with these methods. This subsection will try to clarify the relationship between various methods found in the literature and leave the possibility of incorporating the proposed EM algorithm into the other clustered regression models for future investigation. \par

The $X$ and $y$ covariate can be related through user specified parametric joint distributions. This approaches are called \textit{Cluster Weighted Model} (CWM) with a long line of works development from \citep{RN44,RN45,RN40,RN30} and related works therein. The main strength of CWM over CLR is the the flexibility to perform fine-grain model control through parametric tuning of the data distribution. \citep{RN30} designed several cluster separation criteria to suppress over-fitting tendency of CLR and improve interpretability of the models. However, compared to CLR, CWM is computationally more expensive and more complex to automate when applied in discovering potential hidden relationships in data. High-dimensional dataset containing many clusters will especially accentuate this drawback of CWM. Other interesting developments include \citep{RN32} , which generalizes CLR to perform clusterwise functional regression, and  \citep{RN28}, which performs fuzzy variable prediction instead of scalar prediction.  \par

\section{Quality Metrics and Algorithm}     \label{sec: quality and algorithm}

\subsection{Resolvability}                 \label{sec: resolvability}

This section propose the concept of \textit{resolvability} in the context of clustered regression problems. CLR formulation forces multiple regression functions to a set of data. Under certain circumstances, overfitting and inference of non-existent relationships between $X$ and $y$ will occur. \citep{RN6} pointed out that both clustering and regression can reduce model prediction variance. They argued that CLR models overfits when too many clusters are being used, reducing CLR to a variant of nearest neighbor clustering with the regression within each cluster offering no useful insight into the structure of the data. The Resolvibility Index $R$ is proposed in this section to provide a quantitative indicator to guard against this type of over-fitting. \par

The set of points $\mathbf{D}=(X,y)$ reside in a $p+1$-dimensional space. Each cluster can be approximated by the hyperplane described by equation \ref{equ1}. Let's further assume that $\varepsilon_k \sim \mathcal{N}(0, \sigma_k^2)$. A set of CLR clusters is said to be \textit{irresolvable} if the points from one cluster significantly overlap with those from another, to the extent that inferring cluster membership based solely on the coordinates becomes impossible.\par

\begin{figure}[h]  
    \centering
    \includegraphics[width=1.0\textwidth]{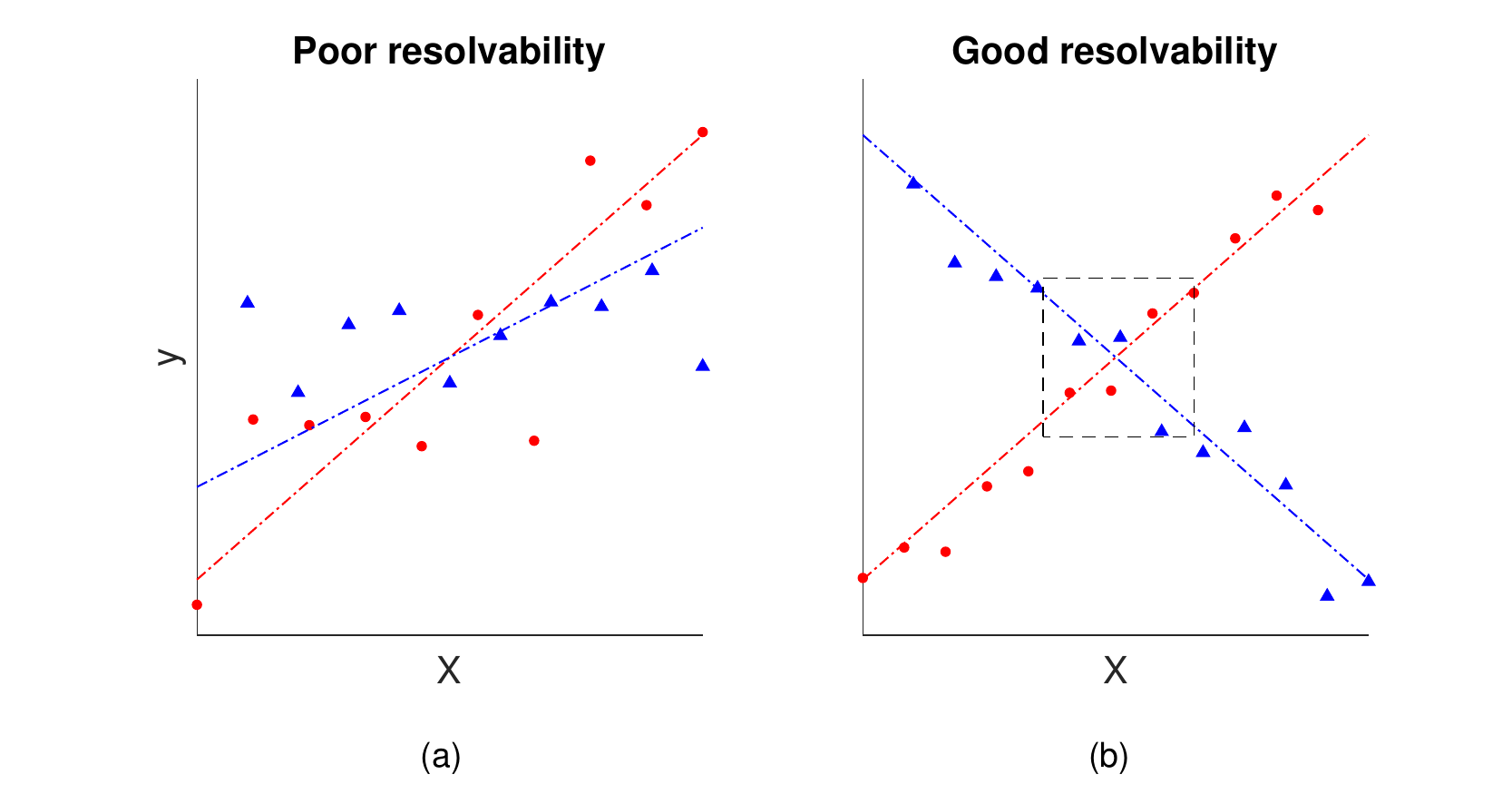}
    \caption{Different levels of resolvability. The dashdot lines represents cluster models; samples, denoted by \textcolor{blue}{$\times$} and \textcolor{red}{$\bullet$}, are generated by the two clusters models perturbed by noise.  (a) Poor resolvability - clusters membership of each point is ambiguous if all differences in markers are removed (b) Acceptable resolvability - The resolvability inside the doted rectangular box is poor but most data point can be assigned cluster membership with good confidence even after all differences in markers are removed. To maintain focus on the qualitative aspects of the illustration, the generative equations have been omitted. } 
    \label{fig: resolvability}
\end{figure}

Figure \ref{fig: resolvability} compares good and poor resolvability visually using two hypothetical one-dimensional examples. Figure \ref{fig: resolvability}(a) illustrates a problem with poor resolvability. It is almost impossible to assign cluster memberships based on the coordinate $(X,y)$ alone even with the knowledge of the ``true model'' $\beta_k$ and $\sigma_k$. The resolvability sets the informational limit on the inverse inference of cluster membership. Figure \ref{fig: resolvability}(b), on the other hand, shows that if the modeling uncertainty and the geometric relationship between clusters hyperplanes satisfy favorable conditions, the CLR problem is resolvable. The dashed-line rectangular box at the centre of Figure \ref{fig: resolvability}(b) marks out a region of relatively poor resolvability amidst a dataset with generally good resolvability outside the box. It follows that smaller volume of the rectangular box implies a better resolvability. \par

These observations inspires the definition the Resolvibility Index $R$ as the complement of irresolvability $R = 1- Q$. The degrees of irresolvability $Q$ can be defined by, assuming the knowledge of model parameters {$\beta_k, \sigma_k$}, the overlapping volume between the clusters. Equation \ref{equ:rho} defines $\rho(\tilde{X},y;\beta,\sigma)$ to measures the intensity of overlap at every coordinate point; $\tilde{X}$ and $y$ in this equation are not the sample but coordinate of the abstract sample space. From this definition, $\rho \approx 0$ leading to $R\approx1$ when there is no overlap between clusters. The normalization constant $Z$ is introduced to keep $0\leq Q\leq1$. \par
\begin{equation}       \label{equ:rho}
    \rho(\tilde{X},y;\beta,\sigma) = Z^{-1} \prod_{k=1}^K \frac{1}{\sqrt{2\pi\sigma_k}} e^{-\frac{1}{2}(\frac{y-\tilde{X}\beta_k}{\sigma_k})^2}
\end{equation}

$\rho(\tilde{X},y;\beta,\sigma)$ has to be integrated over the full space of $\tilde{X}$ and $y$ to assess the how much of nonzero (overlap) contribute to the overall volume of the cluster. Marginalize $\rho(\tilde{X},y;\beta,\sigma)$ over all possible $y$ leads to $G(\tilde{X})$ in equation \ref{equ:G(X)}. Integrating $G(X)$ weighted by $f(X)$, the density distribution of $X$ over the full domain will gives $Q$ as shown in equation \ref{equ:Q}. $Q$ measures the overlap between the clusters, which is what we are looking for. Appendix A provides the derivation and demonstrate, through a few special cases, that $Q$ behaves as expected in measuring the level of difficulty to resolve data into clusters. We have also shown that the expression given in the equation $Z = K^{-\frac{1}{2}}(2\pi s^2)^{-\frac{K-1}{2}}$ where $s = \prod_{k=1}^K \sigma_k^{1/K}$ produces a sensible normalization constant $Z$. Using this choice of $Z$, merging Equation \ref{equ:G(X)} and \ref{equ:Q} leads to Equation \ref{equ:R_explicit} which gives an explicit expression to compute $R$.\par

\begin{equation}        \label{equ:G(X)}
    G(X) = \int_{-\infty}^{\infty}  \rho(X,y,;\beta,\sigma) dy
\end{equation}
\begin{equation}        \label{equ:Q}
    Q = \int_{X} G(X)f(X) dX
\end{equation}

\begin{equation}        \label{equ:R_explicit}
    R = 1 - \frac{\sqrt{\frac{K}{2\pi}}}{\prod_{k=1}^{K}\sigma_k^{1/K}} \int_{X} \int_{-\infty}^{\infty} e^{-\frac{1}{2} \sum_{k} (\frac{y-\tilde{X}\beta_k}{\sigma_k})^2 } f(X) dy dX
\end{equation}

Equation \ref{equ:R_explicit} may look intimidating at the first glance, and the dependence on the unknown distribution $f(X)$ daunting. It should be pointed out that $y$ in Equation \ref{equ:R_explicit} is to be marginalized away; while the integral of $f(X)$ is, by definition, encoded implicitly in the averaging operation of the empirical observations. The approximation of $R$ using discrete empirical observations is given by Equation \ref{equ:R_discrete}. \par

\begin{equation}
    \label{equ:R_discrete}
    R \approx 1 - \sqrt{\frac{K}{\sum_{k=1}^{K}\sigma_k^{-2}}} \frac{\sum_{l=1}^{N} exp \Biggl\{-\frac{1}{2} \sum_{k=1}^{K}  (\frac{\tilde{X_l}^T \beta_k}{\sigma_k})^2 + \frac{1}{2} \frac{ \left[ \sum_{k=1}^{K} \sigma_k^{-2} \tilde{X_l}^T \beta_k \right]^2}{\sum_{k=1}^{K}\sigma_k^{-2}}  \Biggl\}}{N \prod_{k=1}^{K}\sigma_k^{1/K}} 
\end{equation}

The value of $R$ ranges from 0, indicating completely indistinguishable clusters, to 1, indicating clearly separable clusters with no ambiguity. It serves as a conceptual analogue of the Adjusted Rand Index \citep{Rand01121971} for continuous space. The value of $R$  can be used to assess the credibility of the CLR model. The acceptable value depends on the specific application context and usage conventions, much like the interpretation of correlation coefficients. The resolvability index $R$ in equation \ref{equ:R_discrete} is operationally more useful than Hennig’s identifiability criteria \citep{RN1} because:
\begin{enumerate}
    \item   Very simple to evaluate. 
    \item   Only make use of the CLR model parameters $\beta_k$ and $\sigma_k$ and the training data $X$.
    \item   $R$ can be used as a quality indicator of a CLR model. 
\end{enumerate}

The consequence that follows from point 2 and 3 above is particularly important. The evaluation of $R$ according to equation \ref{equ:R_discrete} can use the estimated CLR model parameters because the ``true'' values are almost always unknown in real-life data, this quality indicator will be very useful in sense-checking the validity of the model. A low $R$ value either indicates that a CLR model over-fitted or the problem is fundamentally irresolvable; in either case one should view the CLR model with large dose of skepticism. The computational simplicity in evaluating $R$ allows it to be integrated into the main loop of EM algorithm to assist the search of a better models. A high $R$ is indicative of at least one cluster is resolvable, this characteristics sensitivity to potential subset clustering is an advantages in many situations. \par

\subsection{Making Predictions: X Predictability}                 \label{sec: making predictions}
The utilization of CLR involves two primary goals: (a) identify cluster membership for each training point and identify their within-cluster regression vectors that contribute to explain the response variable, and (b) establishing response predictive capabilities for new not-yet-seen observations. The realization of the first goal (analysis) is attained as soon as the CLR model is generated. This is the traditional finite mixture model analysis of data. However, the feasibility of accomplishing the second objective is contingent upon the dataset characteristics. This is normally related to the deployment of CLR model in science, engineering, business or any other endeavors that requires the use of model to support decision making. A lot of confusion arises in the use of CLR models due to the conflation of these two different goals. To differentiate between problems that allow the user to achieve both goals from the problems that only allow the first goal but not the second, we introduce the concept of \textit{X-Predictability}(XP). A CLR problem is said to to XP if given the independent variable $X$, without the knowledge of the corresponding $y$, it is possible to assign the cluster membership with high confidence and subsequent apply the appropriate cluster-specific regression model for prediction. XP-ness adds nuances to the discussions that involve the use of CLR models. An XP dataset and the CLR model that derived from it can support both analysis and prediction tasks; while a non-XP dataset does not support prediction tasks. Case II and III in Appendix A are one-dimensional examples of XP and non-XP dataset respectively. It should be emphasis that the XP-ness is not describing the quality of CLR model, it is describing the dataset characteristics arise from the separability of the clusters. When the clusters do not overlap, there is little ambiguity on which regression function to deploy for prediction. On the other hand, a non-XP dataset would see the conflict of multiple regression functions on large section of the input domain $X$. This conflict between regression functions is the main reason prediction task from CLR is challenging, especially when the regression vectors of the overlapping clusters are pointing at opposite directions leading to completely opposite predictions. \par

The degrees of XP-ness reduces with the increase proportion of input domain overlap between different clusters. It should be obvious at this point that XP is intimately related to the resolvability index $R$ because both of these concepts are related to the degree of overlap between clusters. All XP datasets have very high $R \approx $1. This has been demonstrated in special case II of Appendix A. However, the reverse is not necessary true as demonstrated in special cases in the special case III of Appendix A. Therefore we can say that high $R$ is the necessary but not sufficient condition to declare XP. On the other hand, low $R$ value rules out prediction as an appropriate use of the CLR model. The example in the next subsection demonstrate and discuss the use of XP in CLR prediction tasks.\par

\subsubsection{Making Prediction}     \label{sec: making predictions results}

In the literature, the methods of making prediction from CLR can be partitioned broadly into two categories. The first category is to coerce membership to the cluster with the highest likelihood score and make prediction using the regression vector of the selected cluster. The cluster assignment is often estimated from the \textit{k-nearest neighbor} or other measures of similarity. The second category of methods generates prediction from all clusters and derive the final prediction from the weighted average of these predictions. The weights are often related to the likelihood of cluster belonging. This likelihood can also come from \textit{k-neaerest neighbor} or the distribution modeling of $X$ for each cluster like what is being done in CWM. Under the second category, one can also train a nonlinear link function $g(\cdot)$  to perform a secondary regression on the CLR predictions to the target value such that $g(X\beta) = y$. This approach is intimately related to the generalized linear regression \citep{RN35} and is achieving the same or better RMSE performance without the hassle of cluster membership analysis and explicit weight assignment. \par

\begin{figure}[h]
    \begin{minipage}{0.5\textwidth}
        \includegraphics[width=1\textwidth]{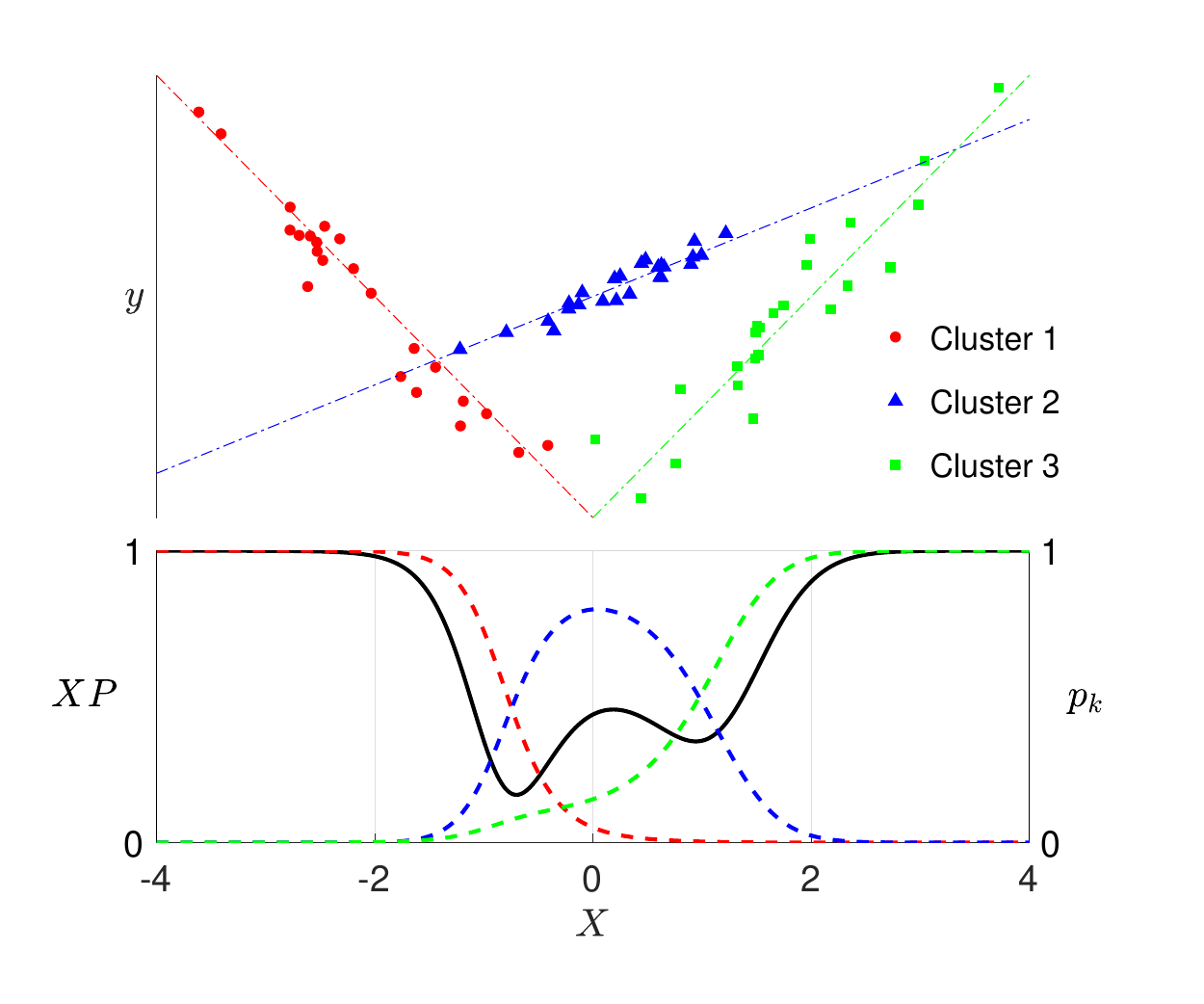}
    \end{minipage}
    \begin{minipage}[t]{0.2\textwidth}
        \tiny %\small
        \vfill
        \begin{tabular}{|r|r|r|r|c|}
        \hline
        $X$ & $\hat{y}_1$ ($p_1$)& $\hat{y}_2$ ($p_2$)&  $\hat{y}_3$ ($p_3$)& $XP$ \\ \hline
        -3 &  1.50(100\%) & 0.40(0\%) & -1.50(0\%) & 1.00 \\ %\hline
        -1 & 0.50(53\%) & 0.80(36\%)& -0.50(11\%) & 0.20 \\ %\hline
         0 & 0.00(4\%)& 1.00(77\%)&  0.00(19\%) & 0.41 \\ %\hline
         3 & -1.50(0\%) & 1.60(0\%)&  1.50(100\%) & 0.99 \\ \hline
        
    \end{tabular}
    \end{minipage}
    \caption{The plot on top shows the point distributions $(X,y)$, the three cluster are $X_1 \sim \mathcal{N}(-2,1), X_2 \sim \mathcal{N}(\frac{1}{4},0.64), X_3 \sim \mathcal{N}(\frac{3}{2},1); y_1 = -\frac{1}{2} X_1, y_2 = 1 + \frac{1}{5} X_2, y_3 = \frac{1}{2} X_3$. On the same horizontal axis $X$ on the plot at the bottom, the solid curve shows how $XP$ changes with the region of $X$; and the dashed lines, the probability of cluster membership. Table on the right: the values of four representative points have been tabulated.} 
    \label{fig: Prediction}    
\end{figure}

The methods in the second category almost always achieve better RMSE performance than those in the first category. This phenomenon can be explained using the example illustrated in Figure \ref{fig: Prediction}. Figure \ref{fig: Prediction} shows a three-cluster, one-dimensional CLR problem. The resolvability of this problem is relatively high due to the low proportion of overlap volume in the two-dimensional $(X, y)$ space. High resolvability means that the correct CLR model can be easily recovered from the observations $(X, y)$. However, when making predictions using only the predictor $X$ and the CLR model—without the corresponding $y$ — the separation between clusters disappears over a large portion of the predictor domain. In this example, there is little ambiguity when $\lvert X \rvert > 2$, as coercing cluster membership or using weighted prediction yields similar results. However, when $\lvert X \rvert \leq 2$, the domains overlap, resulting in comparable probabilities for each cluster. Coercing cluster membership in this region amounts to random assignment, leading to predictions with much higher variance than expected. In contrast, the weighted prediction method attempts to produce a compromise, aiming to reduce the RMSE in these overlapping regions. \par

Although weighted prediction is better in reducing RMSE, this approach produces predictions that do not align with any specific cluster model. For example, a weighted prediction at $X=0$ in Figure \ref{fig: Prediction} would land the predicted point at the empty space away from all three clusters. This caveat in the CLR prediction, if not explicitly declared, could mislead the decision maker. The root cause of this contradiction lies in the desire to present a single scalar prediction, which is at odds with the purpose of performing cluster-wise analysis. A $K$-class CLR model naturally generates $K$ different predictions, and the acceptability of these predictions can be moderated by their associated probabilities. The class membership weight vector $w$ generated as a byproduct of the CLR model construction enables the inference of the probability density function of $X$ for each cluster, illustrated by the dashed lines in Figure \ref{fig: Prediction}. The probability $p_k$ that a point $x \in X$ belongs to cluster $k$ can be inferred from the normalized relative likelihood of these probability density functions at point $x$. The predictions and probabilities for four representative points are tabulated in Figure \ref{fig: Prediction}. \par

To assist with decision-making, we propose quantifying XP using equation \ref{equ: T}, adapted from the definition of normalized multiclass entropy. Under this definition, reducing the prediction to a scalar poses no problem when $XP \approx 1$. However, when $XP$ is small—such as in the cases of $X = -1$ and $X = 0$ in Figure \ref{fig: Prediction}—the one-to-many mapping of CLR cannot be justifiably reduced to a one-to-one mapping without losing critical information. In general, implicitly converting probabilistic results to a single scalar prediction should not be the standard practice. The reduction of CLR predictions to a scalar should be left to the user, who can take into account XP, the probabilities generated by the CLR model, and any other relevant contextual information. \par

\begin{equation} \label{equ: T}
    XP = 1 + \frac{\sum_{k=1}^{K}{p_k log(p_k) }}{log(K)}
\end{equation}
 
A low $XP$ value may indicate a poorly formulated regression problem, where the predictors do not provide enough information for accurate predictions. In such cases, it is more effective to focus resources on improving data sources to enhance regression predictions or acknowledge that CLR predictions is unreliable for the given problem formulation, rather than tweaking computational models to reduce RMSE.

\subsection{Incremental Seeded Expectation Maximization Algorithm}             \label{sec: isem}

The revised EM algorithm proposed in this section will be refered to as the \textit{Incremental Seeded Expectation Maximization Algorithm} EM$_{is}$ (Algorithm \ref{alg: EMis}) to distinguish it from the standard EM algorithm (Algorithm \ref{alg: EM}). EM$_{is}$ differs from EM in adding the blocks 2.5-2.8 (Elite Recombination Block) and 2.14-2.15 (Cluster Revival Block) to generate new $\beta_k$. The reason to introduce these blocks is to help the algorithm escapes local minima. \par

\begin{minipage}{\textwidth}
\begin{algorithm}[H]
\caption{Proposed EM$_{is}$ computational flow}\label{alg: EMis}
\KwIn{$\tilde{X}$, $y$} 
\KwOut{$w$, $\hat{\beta}$, $\hat{\sigma}$} 

$\hat{\beta} = [\hat{\beta_1},...,\hat{\beta_K}] $ \\
$\hat{\sigma} = [\hat{\sigma_1},...,\hat{\sigma_K}] $ \\
Initial guess $w$, $\hat{\beta_k}$, $\hat{\sigma_k}$ for $k={1..K}$ \\
\While{$\neg \textnormal{EarlyTerminate} \wedge (\textnormal{PerturbCount} > 0) \wedge (i < \textnormal{MaxLoop})$}{
    \tikzmark{start1}
    \If{$\textnormal{Converged} \wedge (\textnormal{PerturbCount} > 0)$}{ 
        $\textnormal{New} \hat{\beta_k} \textnormal{combine and perturb best models}$ \\
        $\textnormal{PerturbCount} \leftarrow \textnormal{PerturbCount}-1$     
    }
    \tikzmark{end1}
    $\textnormal{Evaluate fitting error} \varepsilon = y-\tilde{X}\hat{\beta}$ \\
    $\textnormal{Re-estimate scale parameter} \hat{\sigma_k}$ \\ 
    $z^2 = \varepsilon^2/\hat{\sigma}^2$ \\
    $w \leftarrow \phi(z)$ \\
    $\textnormal{Re-weight data using the chosen robust estimator}$\\
    $\textnormal{Estimate the new } \hat{\hat{\beta_k}} \textnormal{using the new } w$ \\
    $\textnormal{Momentum update} \hat{\beta_k} \leftarrow \zeta\hat{\beta_k} + (1-\zeta) \hat{\hat{\beta_k}}$ \\
    \tikzmark{start2}
    \If{\textnormal{cluster collapsed}}{        
        \parbox{4cm}{$\textnormal{generates new } \hat{\beta_k}$ for collapsed clusters}         
        }
    \tikzmark{end2}
    $\textnormal{Keep track of the best models in Elite}$ \\
    $i \leftarrow i+1$ 
}
\end{algorithm}

\begin{tikzpicture}[remember picture,overlay]
  \draw[decorate,decoration={brace,mirror,amplitude=6pt},thick]
    ($(pic cs:end1)+(8cm,0.5cm)$) -- ($(pic cs:start1)+(8cm,0.2cm)$)
    node[midway,right=4pt]{\parbox{2.5cm}{\centering \textnormal{Elite Recombination Block}}};
\end{tikzpicture}
\begin{tikzpicture}[remember picture,overlay]
  \draw[decorate,decoration={brace,mirror,amplitude=6pt},thick]
    ($(pic cs:end2)+(6cm,0.5cm)$) -- ($(pic cs:start2)+(6cm,0.2cm)$)
    node[midway,right=4pt]{\small \textnormal{Cluster Revival Block}};
\end{tikzpicture}
\end{minipage}

\paragraph{Elite Recombination Block} EM$_{is}$, over the iterations, will keep a small set of elites consists of good solutions with small regression error. When the convergence criteria have been met, the Elite Recombination Blocks will attempt to recombine or perturb these elite solutions, restart the EM loop with these new initial solutions. This arrangement allows the EM loop to explore different partial solutions and combine them to escape the local minimum quickly. This recombination block can be viewed as an improvement over the traditional multi-start procedure. It is an improvement because the starting solution is informed and build on top of the best-to-date solutions generated by previous iterations instead of restarting the search blindly. The elite recombination pseudocode is available in Appendix C. \par

\begin{figure}[h]   
    \centering
    \includegraphics[width=0.7\textwidth]{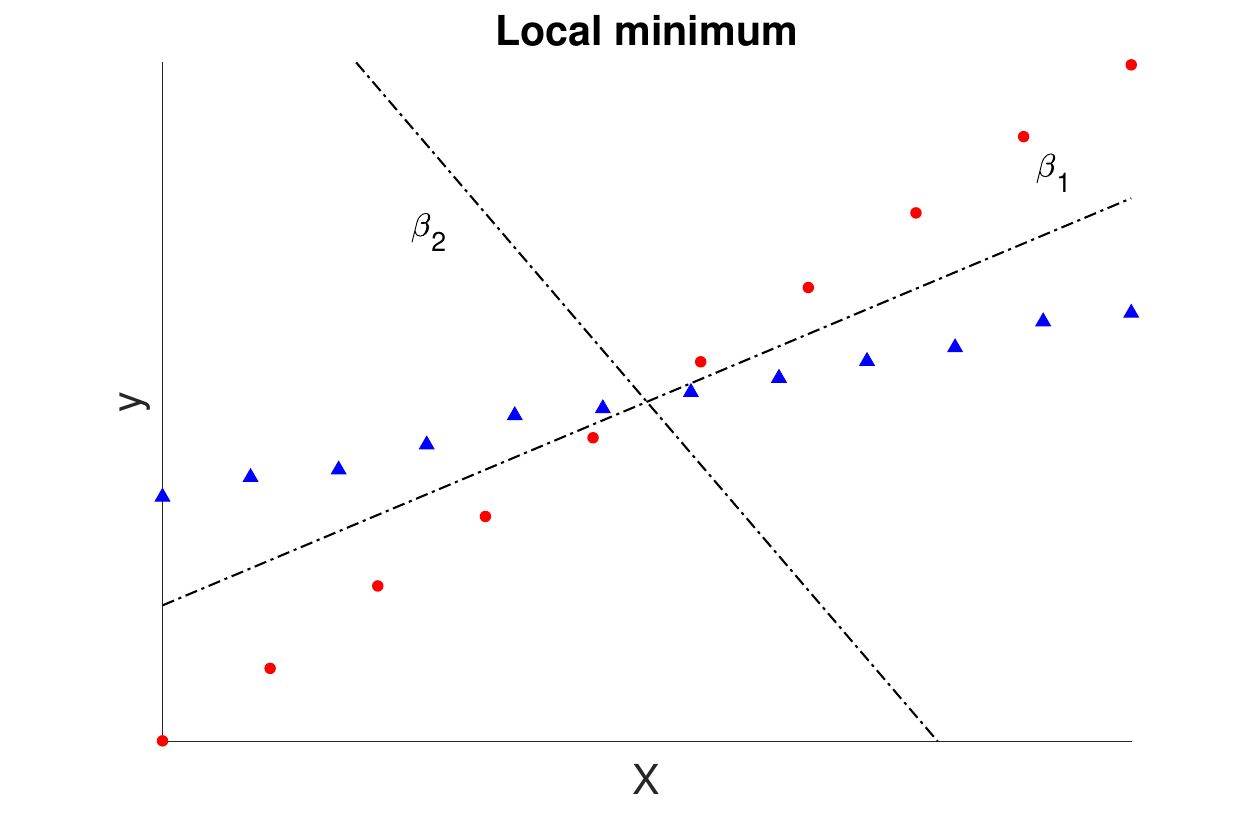}
    \caption{One-dimensional two-cluster example of local minimum. Cluster 1 ($\beta_1$) attracts all data points due to smaller orthogonal distance while cluster 2 ($\beta_2$) collapses into an empty set. Standard EM algorithm is not able to escape this configuration.}
    \label{fig: local minimum}
\end{figure}

\paragraph{Cluster Revival Block} EM$_{is}$ triggers a regression vector re-initialization in Cluster Revival Block when a cluster collapse is detected in Step 2.14. The collapse of weak cluster, as illustrate by Figure \ref{fig: local minimum} using a one-dimensional example, where all members are absorbed by stronger clusters is the most common form of local minimum.  In this example, Cluster 1 has a regression vector $\beta_1$ that is closest to all data points; consequently, all points are assigned to Cluster 1, and Cluster 2 collapses into an empty set. This configuration is suboptimal but extremely stable, making it difficult to escape. Even if $\beta_1$ is perturbed toward the "correct" value, the solution is likely to revert to the same local minimum  shown in Figure \ref{fig: local minimum} unless $\beta_2$ is simultaneously near its true value. The probability of all $\beta_k$ vectors being close to their true values simultaneously decreases geometrically with increasing $K$. Although there is no theoretical reason to believe that all local minima are accompanied by cluster collapse, an overwhelming number of local minima encountered in practice do appear as collapse of at least one clusters. This is most likely due to stability of this particular configuration under random perturbation. The result section shows that EM$_{is}$ designed to address this particular form of local minimum drastically improves the probability of convergence to the correct solutions.  \par

The Cluster Revival Block allows partial re-initialization inside the EM loop when the need arises, lowering the stake of needing to get the first initial solutions correct. This re-initializing within the main EM$_{is}$ loop offers two significant advantages over a initialization before and outside of the main loop in the standard EM flow. First, only the data absorbed by the super-cluster will be used to re-initialize the new regression vectors, interference of irrelevant data from other clusters are being minimized. Second, even though $\beta_1$ in Figure \ref{fig: local minimum} is not the true solution, it is linearly dependent on the true solutions. This additional constrain will reduce the search space of the solution. However, there is no fool-proof indicator to identify a super-cluster reliably. The target cluster can be selected randomly with the likelihood of being selected proportional to the size of the cluster. This approach is based on the heuristic that a super-cluster that inadvertently absorbs multiple clusters is likely to contain larger fraction of the population. \par
 
 In this paper, the cluster collapse condition will be triggered if a cluster membership falls under 10\% of the total dataset size. This seemingly arbitrary threshold value can be justified on the ground that it  does not prevent the true clusters and its regression vectors to be discovered. For example, in the case of extreme cluster size imbalance, e.g. 99\% vs 1\%, the regression vector inferred for the dominant cluster is almost certainly correct while EM$_{is}$ will keep re-estimating new $\beta_k$ for the 1\% cluster and the best solution will be recorded in step 2.17. A more sophisticate and efficient triggering condition is an open problem for further research.  \par
 
\subsubsection{Generating new regression vectors $\beta_k$} \label{sec: new beta}

\begin{figure}[h!]  
    \centering
    \includegraphics[width=0.7\textwidth]{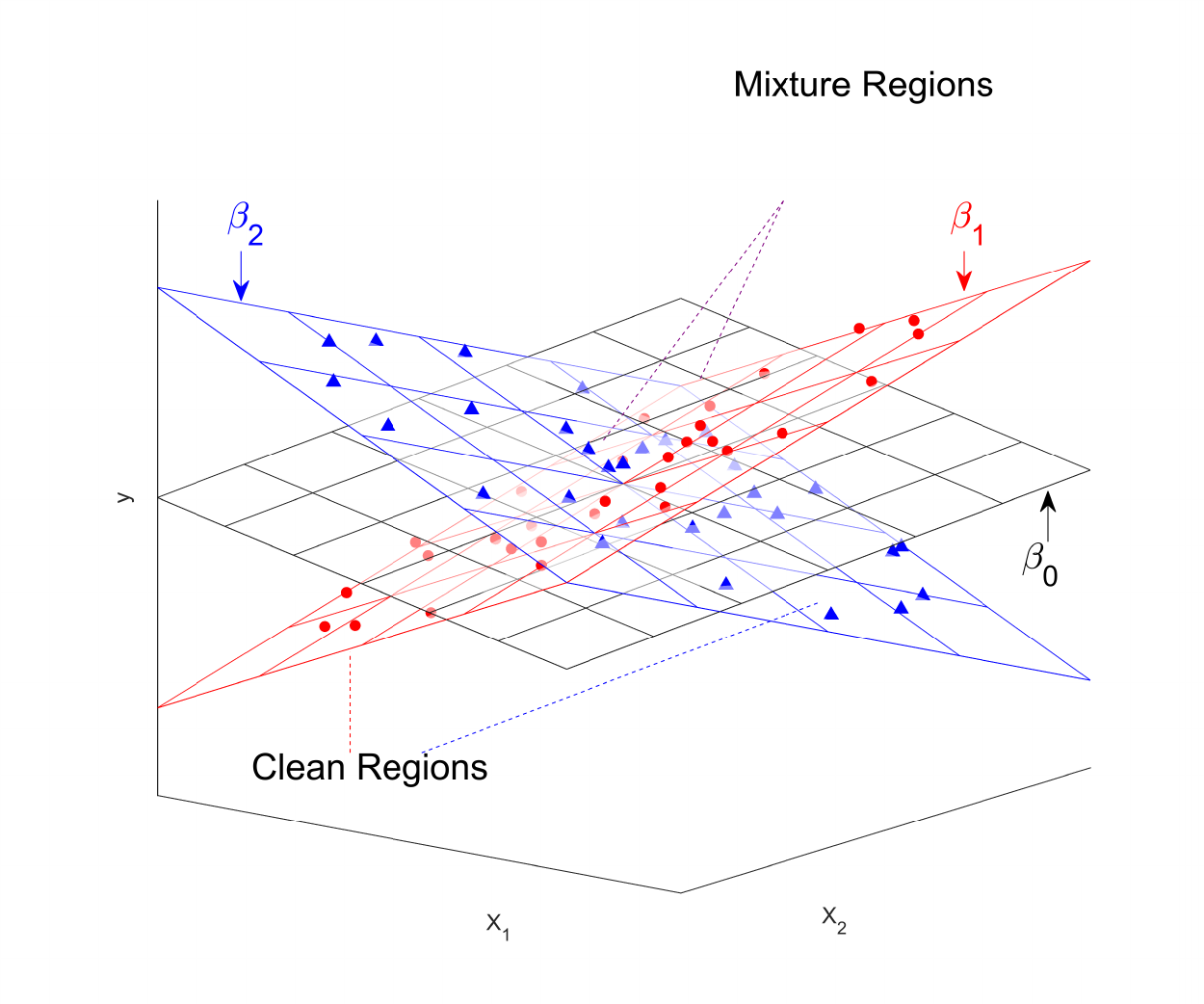}
    \caption{Illustration of point distribution in a supercluster. $\beta_1$ and $\beta_2$ are the correct regression vector while the data points have been trapped in a local minimum $\beta_0$. The data from both clusters will show up in the mixture regions around the intersection of the hyperplanes; while the clean regions away from the intersection contains highly unbalanced representation from the clusters.}
    \label{fig: geometry supercluster}
\end{figure}

The Elite Recombination Block and the Cluster Revival Block are the two places in EM$_{is}$ that will propose new $\beta_k$. Figure \ref{fig: geometry supercluster} shows the point distributions of a supercluster subsuming two smaller clusters when the algorithm enters these two blocks. By plotting the data points as $p+1$-dimensional embedding space, each cluster will form a $p$-dimensional hyperplane. Since the clusters are exactly one dimension lower than the embedding plane, it can be identified using the one-dimensional null space of the hyperplane which is uniquely mapped to the cluster specific regression vector. The objective is to replace the suboptimal solution $\beta_0$ in Figure \ref{fig: geometry supercluster} and the regression vector of the collapsed cluster by the two constitutive components $\beta_1$ and $\beta_2$ that better fits the data. \par

At the first sight, finding the new regerssion vectors is a re-statement of CLR problem and we seem to be back to square one. However, the new problem is likely to be simpler than the original problem by focusing the attention to the sub-population $(X_s,y_s) \in \mathbb{R}^{M \times (p+1)}$ that is already attracted into the supercluster by $\beta_0$. We further restrict the algorithms to generate only two regression vectors at a time. This restriction does not affect our ability to handle supercluster that subsumes more than two clusters because the supercluster can be decomposed incrementally by splitting it into two at a time. On the other hand, limiting our attention to resolve two clusters significantly reduces the algorithmic complexity by circumventing the need to estimate the number of subsumed clusters. This section will propose Edge-Point K-flat (Algorithms \ref{alg: EdgePoint}) and Center-Point Splitting (Algorithms \ref{alg: CenterPoint}) to generate promising new $\beta_k$ from the available information. Both these algorithms are rooted in the geometry of the cluster subspace shown in Figure \ref{fig: geometry supercluster} but take advantage of different geometrical characteristics to achieve the goal. \par

\paragraph{Edge-Point K-flat Algorithm (Algorithm \ref{alg: EdgePoint})} K-flat subspace clustering algorithms were introduced by \citep{RN20,RN43,RN50}, where the cluster hyperplanes are referred to as K-flats. Given selected samples from a cluster, the regression vector can be estimated by identifying the smallest eigen-direction of the sample covariance. This approach works as long as the regression uncertainty is smaller than the uncertainty along the other eigen-directions of $X_s$. We denote the K-flat procedure of finding the regression vector $\alpha$ from the sample point set ${P}$ by $\alpha \leftarrow \mathscr{H}(P)$. The primary challenge in this method is obtaining sufficient clean samples from the same cluster to perform eigen-decomposition without encountering numerical singularity.

The principal component projection $\{P,\alpha_0\} = \mathscr{T}(Z,\beta_0)$ in step 3.1 maps the points $Z$ to $P$ and the regression vector $\beta_0$ to $\alpha_0$ via the same transformation. This pre-processing step reduces the negative impact of correlated predictors on the algorithm performance. Appendix B provides a details of the transformation $\mathscr{T}$, as well as the corresponding inverse mapping $\mathscr{S}$ used in step 3.20. Steps 3.2-3.3 identify the group of points located in the clean region, as illustrated in Figure \ref{fig: geometry supercluster}. This is done by shortlisting points that are furthest from the $\beta_0$ hyperplane. The ``straightness'' of the linear regression models implies that the local resolvability will increase with increase orthogonal distance from the group-average hyperplane $\beta_0$. Figure \ref{fig: geometry supercluster} shows that, mixture region can be far from centroid in high dimensional spacess, the orthogonal distance from the group-average hyperplane is a better pointer to find the clean regions.  Therefore only points at the highest $f$ percentile of the orthogonal distance will be included in the shortlist set $\{P_t\}$. Algorithm \ref{alg: EdgePoint} is not sensitive precise value of $f$ because most of the redundant points will be removed in step 3.12. In this paper, percentile $f$ between 5 and 15 will be generated random every time Algorithm \ref{alg: EdgePoint} is being called. \par

\begin{algorithm}[h!]    \label{alg: EdgePoint}
\caption{Edge-Point K-flat} 
\KwIn{$X_s \in \mathbb{R}^{M \times p }$, $y_s \in \mathbb{R}^{M \times 1}$, $\beta_0 \in \mathbb{R}^{(p+1) \times 1}$}
\KwOut{$\beta_{new}$}
\Parameters{$f$, $k_{nn}$, $\Xi$}
\Definitions{$Z = [X_s,y_s]$}

    $\{P,\alpha_0\} = \mathscr{T}(Z,\beta_0)$ \\
    Calculate orthogonal distance: $L = \vert P \check{\alpha}_0 \vert $ \\
    $ \mathcal{P}_f = \{ P_{i,:} \mid L_{i} $ in highest $f$ percentile of $L \}$ \\
    \While{$\mathcal{P}_f \neq \emptyset$}
    {   Select one $P_{s_1} \in \mathcal{P}_f$ \\
        $\mathcal{P}_{s_1} = k_{nn}$ nearest neighbors of $P_{s_1}$ \\
        $\alpha_1 \leftarrow \mathscr{H}(\mathcal{P}_{s_1}) $, $\qquad s_1 =$ std $(\mathcal{P}_{s_1} \check{\alpha}_1)$ \\    
        Find $P_{s_2} \in \{P_{i,:}\}$ such that $ \vert \tilde{P}_{s_2} \alpha_1 \vert \geq \vert \tilde{P}_{i,:} \alpha_1 \vert \quad \forall \tilde{P}_{i,:}$ \\
        $\mathcal{P}_{s_2} = k_{nn}$ nearest neighbors of $P_{s_2}$ \\
        $\alpha_2 \leftarrow \mathscr{H}(\mathcal{P}_{s_2}) $, $\qquad s_2 =$ std $(\mathcal{P}_{s_2} \check{\alpha}_2)$ \\
        $\mathcal{P}_r = \{ P_i \in \mathcal{P}_f \mid \underset{k}{\min} \vert \frac{\tilde{P}_{i,:} \alpha_k}{s_k} \vert < \Xi \} $   \\
        $\mathcal{P}_f \leftarrow \mathcal{P}_f \setminus \{ \mathcal{P}_r, \mathcal{P}_{s_1}, \mathcal{P}_{s_2}\} $      \\
        $ E = \sum_{i} \underset{k}{\min} \vert \tilde{P}_{i,:} \alpha_k \vert $ \\
        \If{$E < E_{best}$}{
            $E_{best} \leftarrow E$ \\
            $\alpha = [\alpha_1,\alpha_2] $ \\
            %$ C_{k}  = \tilde{P}_{i,:}$ such that $ \vert \tilde{P}_{i,:} \alpha_k \vert < \vert \tilde{P}_{i,:} \alpha_{k'} \vert, \quad \forall k' \neq k$ 
        }
    } 
    (Optional) optimization $ \alpha \leftarrow \underset{\alpha}{\arg\min\{E\}} $ \\
    $\beta_{new} = \mathscr{S}(\alpha)$
\end{algorithm}

In the main loop, a random point $P_{s1}$ is selected from the set ${P_t}$, and its $k_{nn}$ nearest neighbors are identified. These local samples are used to estimate the best-fitting K-flat hyperplane in step 3.7. There is no strict rule for selecting the value of $k_{nn}$, except that it should be larger than the dimension of the vector $P$ to avoid singularities in calculating $\mathscr{H}(\mathcal{P}{s_1})$, but not so large as to include points from mixed regions. In practice, setting $k_{nn}$ slightly larger than the dimension of $P$ works well. \par

Next, in step 3.7, the standard deviation $s_1$ of the sample points is estimated based on their distance from the newly found hyperplane $\alpha_1$. Once $\alpha_1$ is determined, the points furthest from it are likely to belong to another cluster, represented by the hyperplane $\alpha_2$. This is a manifestation of the two-cluster assumption. With both $\alpha_1$ and $\alpha_2$, along with their respective uncertainties $s_1$ and $s_2$, the remaining points in the set $\mathcal{P}_f$ that lie within $\Xi$ times the standard uncertainty from these hyperplanes can be assigned to these cluster and removed from further consideration. As soon as a good solution pairing $\alpha_1$ and $\alpha_2$ is found, step 3.12 collapses the set $\mathcal{P}_f$ and the loop terminates quickly. If a poor initial point $P{s1}$ is selected, resulting in a bad estimate, the loop will restart with a different initial point. Step 3.12 removing redundant points is critical to maintain high computational efficiency and making the algorithm insensitive to the value of $f$, $k_{nn}$, and $\Xi$. A value of $\Xi = 3$ is used in this paper.\par

The vector $\alpha$ can be optimized using EM algorithm in the original K-flat formulation, in step 3.19 before returning the main EM$_{is}$ loop. However, it is not worthwhile allocating significant computational resources on this optimization, as the main algorithm EM$_{is}$ only requires new estimate to restart its own EM optimization. Therefore step 3.19 is optional and, if performed, should bee restricted to very few iterations.

\paragraph{Center-Point Splitting Algorithm (Algorithm \ref{alg: CenterPoint})} This procedure will attempts to split the supercluster regression vector $\beta_0$ in Figure \ref{fig: geometry supercluster} into $\beta_{1}$ and $\beta_{2}$ through a displacement vector $\nu$ such that $\beta_{1},\beta_{2} \approx \beta_0 \pm \nu$. The main challenge is to deduce the direction and length of $v$ from the data attracted to the supercluster. Algorithm \ref{alg: CenterPoint} operates in the principal component projected space like Algorithm \ref{alg: EdgePoint}. The principal component projection moves the centroid to be the origin and the eigen-directions are parallel to the cannonical axes. Figure \ref{fig: geometry supercluster} will be helpful to visualize the procedures to be described below. \par

Step 4.2-4.4 finds the orthogonal distance from the plane defined by $\alpha_0$ and their projection onto this plane. If a displacement vector $\gamma v$ exists such that $\alpha_1, \alpha_2 = u_0 \pm \gamma v$, then the group of points that generates $L \approx 0$ (approximately 50 percentile because $L$ is signed distance) will have almost zero spread in the $v$ direction whereas the distribution of points along other directions are not significantly affected. Step 4.5-4.6 of Algorithm \ref{alg: CenterPoint} determines the direction $v$ through eigen-decomposition of the sampled points near the intersection between the two hyperplanes, hence named Center-Point Splitting. The eigenvectors $V$ will have $v$ as the eigenvector associated with the smallest eigenvalue. \par

\begin{algorithm}[h]    \label{alg: CenterPoint}
\caption{Center-Point Splitting} 
\KwIn{$X_s \in \mathbb{R}^{M \times p }$, $y_s \in \mathbb{R}^{M \times 1}$, $\beta_0 \in \mathbb{R}^{(p+1) \times 1}$}
\KwOut{$\beta_{new}$}
\Parameters{Several $\Theta_i = (\Theta_{i}^{low},\Theta_{i}^{hig})$ pairs such that $0 < \Theta_{i}^{low} < 50 < \Theta_{i}^{hig} < 100$}
\Definitions{$Z = [X_s,y_s]$; $\alpha_1(\gamma) = \frac{\check{\alpha}_0 + \gamma v}{\lVert \check{\alpha}_0 + \gamma v \rVert} $; $\alpha_2(\gamma) = \frac{\check{\alpha}_0 - \gamma v}{\lVert \check{\alpha}_0 - \gamma v \rVert} $}

    $\{P,\alpha_0\} = \mathscr{T}(Z,\beta_0)$ \\
    $u_0 \leftarrow \check{\alpha}_0 / \lVert \check{\alpha}_0 \rVert $ \\
    Signed distance $L = P u_0$ \\
    $Q \leftarrow P - L u_0^T$ \\
    $Qs \leftarrow \{Q_{i,:} \mid L_i \in [\Theta_{1}^{low},\Theta_{1}^{hig}]$ percentile of $L \}$ \\
    Generate probing bases $V =$ eigenvector of $Q_s^TQ_s$ \\
    \ForEach {i}{
        $Qs \leftarrow \{Q_{l,:} \mid L_l \in [\Theta_{i}^{low},\Theta_{i}^{hig}]$ percentile of $L \}$ \\
        $\lambda_i \leftarrow diag(V^T Q_s^T Q_s V)$
    }
    $v = $ component of $V$ that produces largest variation in $\lambda_i$ \\
    $\gamma^* = \underset{\gamma}{\arg\min (d^T d)}$ where $d = \min [ P \alpha_1(\gamma), P \alpha_2(\gamma) ]$ \\
    
    $\beta_{new} = \mathscr{S}([\alpha_1(\gamma^*),\alpha_2(\gamma^*)])$
    
\end{algorithm}

Unfortunately, the closer we get to the hyperplane, the smaller the number of eligible samples hence reducing the reliability of the eigen-decompoosition. To address this weakness, block 4.7-4.10 has been added to evaluate $\lambda_i$, the spread of points along all $V$ directions, at different percentile ranges $\Theta_i$. The same basis $V$ generated in step 4.6 are being used inside the loop to avoid unwanted perturbation of the probing bases $V$ due to random point distribution. As the range $\Theta_i$ increases from $(50^-,50^+)$ towards $(0,100)$, $\lambda_i$ increase from zero to the eigenvalue of $P$ in direction of interest; while $\lambda_i$ in all other directions are only perturbed around the eigenvalue. Although denser grid of $\Theta_i$ would generate smoother $\lambda_i$, it was found that smoother $\lambda_i$ does not offer more advantage in finding the correct direction $v$. Therefore, only three ranges $\Theta_1 = (45,55), \Theta_2 = (25,75), \Theta_3 = (5,95)$ are being used in this paper. Step 4.11 select $v$ from $V$ by looking for the component of $\lambda$ that experience largest relative variance. Once $v$ has been found, the optimal length $\gamma^*$ that minimize fitting error $d^T d$ can be found efficiently using bisection method. \par

\paragraph{Summary} The situations that calls for these algorithms necessarily having significantly overlapping predictor $X$ of both clusters, hence the assumption of identical range of $X$ in their derivation. If the two clusters are separated by a large distance, the standard EM iterations would easily identify the clusters at the kmeans clustering step and find the solution. In principle, the reliability of both algorithms are independent of the dimensionality of the input $X$. The Edge-Point K-flat Algorithm and the Center-Point Splitting Algorithm rely on different statistics to resolve a supercluster into its constituent clusters. The Edge-Point K-flat Algorithm rely on having sufficient data points in the clean region; onthe other hand, the Center-Point Splitting Algorithm is more stable when there are sufficient data points near the hyperplane intersection region. These the two algorithms agree most of the time but complement each other depending on the data point distribution. Consequently, EM$_{is}$ will randomly choose from these two algorithms with equal probability of 50\% every time a new proposal of $\beta_k$ is needed.

\section{Methods}       \label{sec: methods}

In the following discussion, the iteration is considered converged if the regression errors of the last $n_{c}=7$ iterations are not monotonically decreasing and that the average relative regression error of $n_{c}$ is below the user defined threshold $T_{c}=10^{-2}$. 

\subsection{Problem Class Characterization}     \label{sec: problem characterization}

Real data has its place in demonstrating the usefulness of an algorithm in real-life scenario but is not ideal to compare the performance of algorithms due to lack of control over various problem characteristics. The main goal of this paper is to compare performance of the proposed EM$_{is}$ algorithm against the standard EM algorithm in the context of CLR problem. Therefore simulated problems with known solutions instead of real problems are being used. We wish to set up the comparison such that the results are statistically significant. All comparisons are achieved via solving automatically and independently generated problems. We adopt the framework proposed by \citep{Kuang} where CLR problems will be characterized by the number of clusters ($K$), dimension of the independent input ($p$), sample size per cluster ($N_k$), angular separation between regression vectors measured by their dot product ($dp$), noise ($\eta$)  and cluster center offsets ($\Delta$). In the following results, we will only focus on the most challenging cases of $\Delta=0$ as explained in Section \ref{sec: challenge}. Other characteristics may be adjusted depending on the needs and will be indicated with the results.\par

The problems are characterized by the sample size per cluster ($N_k$) rather than the total sample size, allowing for easier comparison of performance across problems with varying numbers of clusters ($K$). The noise follows the distribution $\varepsilon_k \sim \mathcal{N}(0, \sigma_k^2)$, where $\sigma_k$ is scaled automatically according to the natural variance of the cluster-specific signal via the relation $\sigma_k = \eta (\tilde{X} \beta_k)$. The parameter $dp$ controls the pairwise dot products between the regression vectors $\beta_k$: a value of $dp \approx 0$ corresponds to mutually orthogonal clusters, while a large $dp$ indicates small angular separation between clusters. This design enables controlled analysis of the impact of angular separation by isolating cases with consistent inter-cluster angles. The qualitative interpretation of $dp$ can then be extended to scenarios involving mixed angular separations. The procedure for generating test datasets with these properties was algorithmically described by \citet{Kuang}, enabling a large number of independent problems to be tested for statistically robust conclusions. Finally, without lost of generality, we further impose independence assumption between the columns of $X$.  Therefore the dimensionality $p$ in the following section should be read as the ``effective dimensionality'' of the independent input components.\par

\subsection{Performance Metrics}    \label{sec: performance metrics}

The root mean square error (RMSE) is a commonly used goodness-of-fit metric in the CLR literature. However, its use will be limited in the following discussions due to interpretive ambiguity. A small RMSE could indicate either a good recovery of the regression vector $\beta$ or overfitting, as explained by \citep{RN6, RN31, RN42}. The primary accuracy metric used in this paper is $ACC$, defined in equation \ref{equ: Acc}, which measures the relative accuracy of the regression vector $\hat{\beta}$ recovered by the algorithm in comparison to the true value $\beta$. The use of $ACC$ requires the columns of $X$ to be independent to ensure the uniqueness of the optimal regression vector. Additionally, because $ACC$ relies on the known true value $\beta$, it is primarily applicable to the analysis of simulated data. Nevertheless, $ACC$ does not suffer from the interpretive ambiguity associated with RMSE. Despite its limitations, Section \ref{sec: resolvability result} demonstrates empirically that $ACC$ is related to resolvability, which can be directly estimated from observed data without needing to know the true values of $\beta$.  \par

\begin{equation}
    \label{equ: Acc}
    ACC = \frac{1}{K} \sum_{k=1}^{K}{ \left[ \max( \: 0, \: 1 - \frac{\Vert \hat{\beta_k}-\beta_k \Vert_2}{\Vert \beta \Vert_2} \:) \right] }
\end{equation}

$ACC$ for a single problem measures how close the estimated regression vector $\hat{\beta}$ is from the true value $\beta$. The average $ACC$, obtained by averaging across many independent instances CLR problems in a same class, indicates the reliability of an algorithm in finding the correct solutions for the given class of problems. From this perspective, the average $ACC$ combines the probability of converging to the true solution and the accuracy of recovering the true regression vector, without the need to define the convergence threshold. By removing the need to define an arbitrary threshold that defines convergence, the average $ACC$ is easier interpret as an indicator of an algorithm's reliability in recovering the true underlying structure of the data.

\section{Results and Discussions}       \label{sec: results}

\subsection{Returning Correct Model } \label{sec: algorithm performance}

\begin{figure}[h]  
    \centering
    \includegraphics[width=1.0\textwidth]{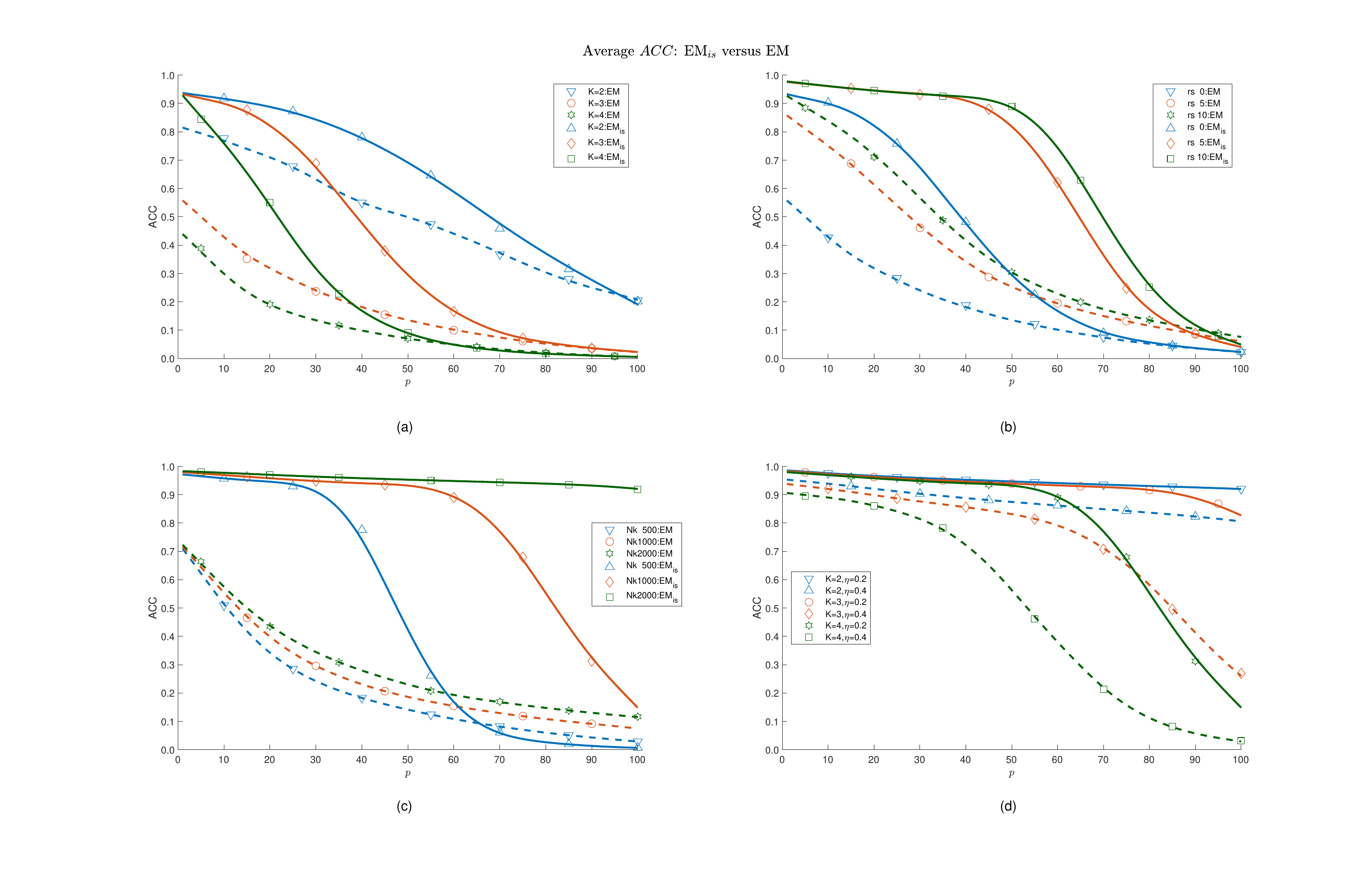}
    \caption{$ACC$ performance comparison between EM and EM$_{is}$. (a) 0 restart, $N_k = 500$, $dp=0.2$, $\eta=0.2$. (b) $K=3$, $N_k =500$, $dp=0.2$, $\eta=0.2$ (c) 10 restarts, $K=4$, $dp=0.2$, $\eta=0.2$. (d) EM$_{is}$, 10 restarts, $N_k=1000$, $dp=0.2$.}  
    \label{fig: ACC vs p}
\end{figure}

Figure \ref{fig: ACC vs p} presents $ACC$ as a function of the input dimension $p$. Each point represents the average $ACC$ over 1,000 independent random problems from the same class, characterized by parameters $K$, $p$, $N_k$, $dp$, and $\eta$ (see Section \ref{sec: problem characterization}). 

From the understanding of cluster resolvability, We expect $ACC$ to deteriorate as $\eta$ and $dp$ increase. To avoid overcrowding the graphs, specific values of $dp=0.2$ and $\eta=0.2$ were selected for the presentation in Figure \ref{fig: ACC vs p} (a)-(c). Figure \ref{fig: ACC vs p} (a) shows that the standard EM algorithm is only reliable for $K=2$ and very low $p$, with its performance rapidly deteriorating as $K$ and $p$ increase. This result helps explain why the published literature on CLR primarily involves low-dimensional datasets. The performance decline with increasing $p$ can be attributed to the exponential growth of the solution search space as the dimension $p$ increases, while the standard EM algorithm lacks mechanisms to recover from local minima in the search process. In contrast, EM$_{is}$ shows significant improvement, maintaining higher $ACC$ for larger values of $K$ and $p$. The observed performance gain shown in Figure \ref{fig: ACC vs p} (a) is due solely to the Cluster Revival Block of EM$_{is}$, because the Elite Recombination Block and multirestart were turned off (0 restarts) during the computation. \par

Figure \ref{fig: ACC vs p} (b) illustrates the $ACC$ improvements across different numbers of restarts for the same class of problems. As explained in Section \ref{sec: isem}, the Elite Recombination Block in EM$_{is}$ provides more directed solution improvements compared to the blind restart approach often used with the standard EM algorithm. It is evident that $ACC$ improvements from multiple restarts taper off quickly as the solution search space expands. In contrast, EM$_{is}$ demonstrates qualitatively superior improvements. For instance, for problem classes with $K=3$ and $p<50$, can be effectively solved using EM$_{is}$ with no more than 10 restarts. \par

Although Figures \ref{fig: ACC vs p} (a) and (b) show that EM$_{is}$ significantly improves $ACC$ performance over the standard EM algorithm, the performance still declines as $p$ increases. This observation may seem to contradict the claim in Section \ref{sec: new beta} that both the Edge-Point K-Flat Algorithm and the Center-Point Splitting Algorithm have weak dependence on $p$. Figure \ref{fig: ACC vs p} (c) clarifies this apparent contradiction by demonstrating that $N_k$ (the sample size per cluster) has a significant impact on the performance of the algorithm. Increasing the sample size restores EM$_{is}$ ability to find the correct solutions. In other words, the performance decline in Figures \ref{fig: ACC vs p} (a) and (b) is caused by a small sample size relative to the problem's dimensionality. The Edge-Point K-Flat and Center-Point Splitting algorithms both rely on eigen-decomposition to find the optimal direction for resolving superclusters into their constituent clusters. In higher-dimensional spaces, more data points are required to numerically stabilize the eigen-decomposition, but both algorithms rely on certain percentiles of the total sample. The number of "useful" data points grows sublinearly with respect to the sample size $N_k$. The most important implication of this result is that CLR modeling of high-dimensional data is inherently data-hungry. Another important inference from Figure \ref{fig: ACC vs p} (c) is the stark qualitative difference in reliability between EM$_{is}$ and traditional EM when the number of cluster $K$ is larger than 2. \par

Figure \ref{fig: ACC vs p} (d) illustrates the impact of noise on the performance of EM$_{is}$. The large-noise data is more difficult to model due to reduced resolvability as explained in Section \ref{sec: resolvability}. When the noise is too large, CLR modeling becomes ill-posed from information perspective. The sharp drop in $ACC$ observed in Figure \ref{fig: ACC vs p} (d) is consistent with this prediction. In general, this sharp drop in $ACC$ is to be expected when EM$_{is}$ reaches the informational limit imposed by the data either due to noise, sample size or any other reasons. \par

Finally, it is important to read the results presented in Figure \ref{fig: ACC vs p} with the understanding that it is the performance under worst case scenario where the predictor $X$ of all the clusters completely overlaps ($\Delta=0$). The dependence on $p$ and $N_k$ will be very significantly milder when the domain of $X$ between clusters do not overlap. The concept of resolvability provides a unified framework to evaluate the difficulty of recovering CLR model in all cases regardless of the value of $\Delta$.

\subsection{Resolvability Index}  \label{sec: resolvability result}

Section \ref{sec: resolvability} introduces the resolvability index $R$ as a metric that measures how amenable a problem is to the CLR analysis. $R$ summarizes the net impact of noise and geometric relationship between clusters without needing to explicitly and individually  estimate these factors.  This section shows that it can also be used, in complementary to RMSE, to assess the quality of the CLR model. This is done by connecting the following two facts: first, a high $ACC$ indicates high probability that the CLR model has recovered a true underlying relationship between $X$ and $y$; second, the estimation of $R$ only uses the observed data and CLR model parameters. By establishing the relationship between $R$ and $ACC$, we can use $R$ as a proxy to gauge the possible range of $ACC$, hence model quality in a real-life problems. \par

In problems involving more than two clusters ($K > 2$), the global resolvability index $R$, which performs well for two-cluster cases, loses its interpretive power due to mixed pairwise resolvability. In such situations, the presence of even one highly resolvable cluster pair can elevate the global resolvability index $R$ close to 1, despite other clusters being poorly separated. Consequently, a scalar $R$ fails to provide a nuanced view of cluster separability when multiple clusters are involved. To address this limitation, it is preferable to examine pairwise resolvability indices $R_p$, which offer a more granular understanding of the separability between each pair of clusters. The dimension of $R_p$ is the binomial coefficient $\binom{K}{2}$, reflecting the number of cluster pairs in a system of $K$ clusters. These pairwise resolvability indices allow for fine-grained control and interpretation of the separability among different cluster pairs. Taking advantage of the fact that the naming order of clusters does not impact the resolvability analysis, $R_p$ can be represented as a vector, with its components ordered from highest resolvability (best-separated pairs) to lowest resolvability (least-separated pairs). This ordering reduces combinatorial complexity through symmetry and facilitates the assessment of the overall distribution of cluster separability, rather than relying on a single $R$ value that could obscure poor separability in certain cluster pairs. This approach provides a more detailed and accurate representation of the overall structure of the clustering problem when $K > 2$. \par

To enable an analysis that links the vector $R_p$ to the scalar metric $ACC$, we define a smooth $\binom{K}{2}$-to-one function $\psi_K(R_p)$ for each $K$-cluster problem, trained to minimize the error $\lVert \psi_K(R_p) - ACC \rVert$. The monotonic ordering of the components in $R_p$ significantly reduces the unnecessary degrees of freedom in $\psi_K$. However, the complexity of $\psi_K$ still increases rapidly with $K$, and training faces the curse of dimensionality. Despite the challenges in expressing $\psi_K$ analytically, the concept remains useful as long as it can approximate $ACC$. The specific form and construction of $\psi_K$ is an interesting research topic but not the primary focus of this section. $\psi_K$ will be realized here using epsilon-insensitive support vector machine regression with a radial basis kernel with automated hyperparameter tuning. \par

\begin{figure}[h]  
    \centering
    \includegraphics[width=1.0\textwidth]{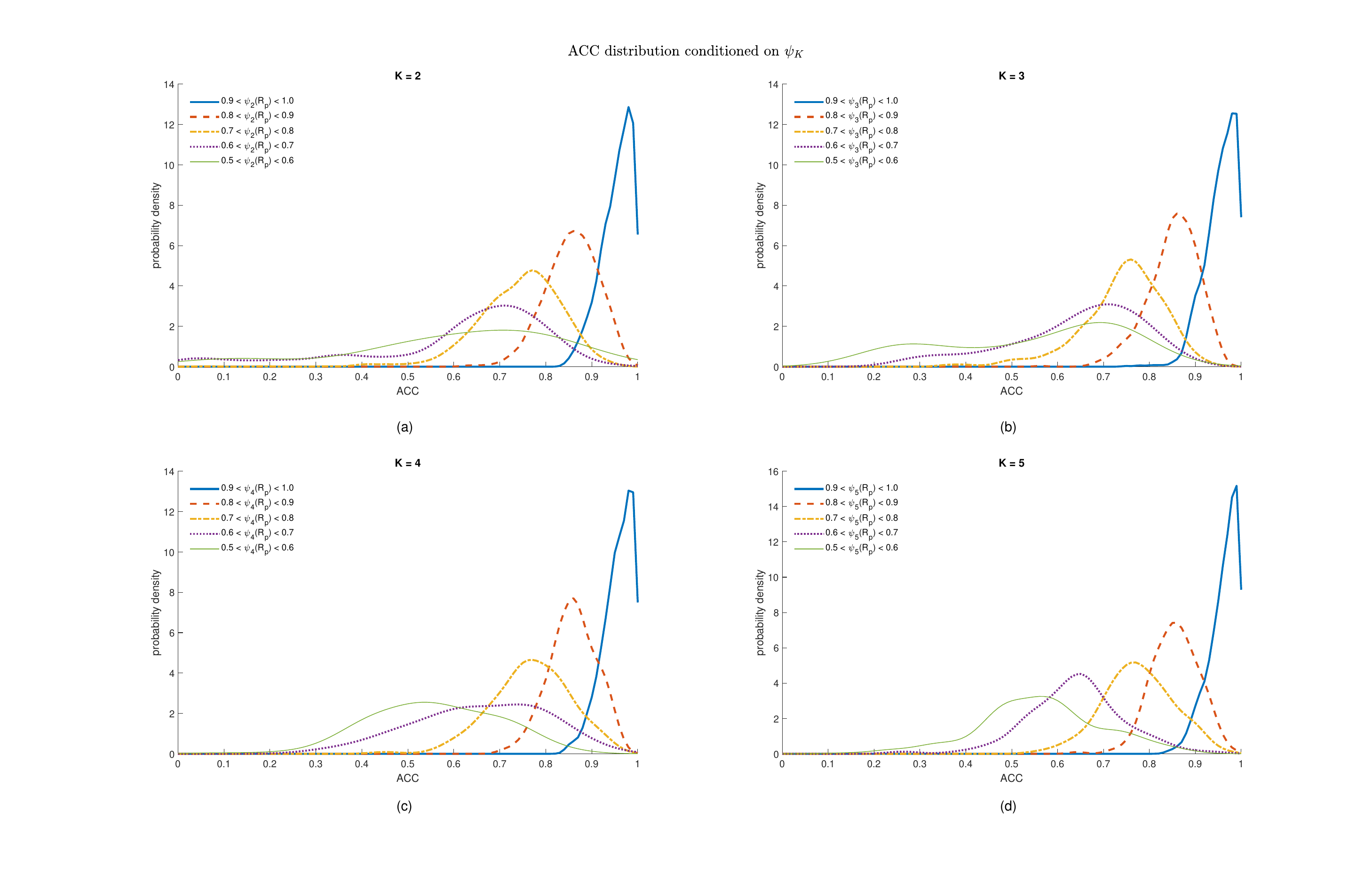}
    \caption{The distribution of $ACC$ as a function of $\psi_K(R_p)$ for different $K$ assuming completely overlapping predictors $X$. High pairwise resolvability index $R_p$ indicates reliable CLR solutions.} 
    \label{fig: R vs ACC}
\end{figure}

Figure \ref{fig: R vs ACC} illustrates the relationship between $ACC$ and $R_p$ through the mapping function $\psi_K(R_p)$. It plots the probability density function of the empirical $ACC$ achieved by EM$_{is}$, conditioned on different ranges of $\psi_K(R_p)$. The probability density function is approximated from the empirical data using univariate gaussian kernel density estimation. The values of $R_p$ used in the calculation of $\psi_K(R_p)$ are based on the empirically estimated values $\hat{\beta_k}$ and $\hat{\sigma_k}$ obtained from EM$_{is}$. A total of 14,000 random problems were generated with characteristics $K \in \{2, 3, 4, 5\}$, $0 \leq dp < 1$, $0 \leq \eta \leq 1$, and $\Delta = 0$, and were solved using EM$_{is}$. Note that $dp$ is no longer required to be scalar in this section. The input dimensionality was controlled within the range $5 < p < 100$, and based on the discussion from the previous subsection, $N_k = 1500$ with 10 restarts was sufficient to solve problems in this class if the formulation is resolvable.\par

The trend in the probability density function indicates that the estimated pairwise resolvability indices, $R_p$, are strongly correlated with $ACC$. However, the uncertainty in the mapping from $R_p$ to $ACC$ increases as the value of $\psi_K(R_p)$ decreases. As a general rule, a high value of $\psi_K(R_p) > 0.8$ is a strong indication that a true solution can be found, whereas a lower value of $\psi_K(R_p) < 0.7$ raises concerns about the resolvability of the problem or that the model overfitted. Although the results are limited to $K \leq 5$, considering the same pattern shows up regardless of the number of clusters, the trend is expected to continue for larger $K$. Compare the probability density of low $\psi_K$ values in Figure \ref{fig: R vs ACC}(d) and (a), increasing $K$ increases the likelihood of encountering problems with poor resolvability.  \par

Two-cluster problem has a particular simple relationship that worth mentioning. $\psi_2$ that generates Figure \ref{fig: R vs ACC}(a) has the form $\psi_2(R_p) = R_p = R$. In other words, one can use simple relationship $E[ACC] = R$ in two-cluster problems. This mapping can be used with the caveat that it is only correct to the expectation of $ACC$ and the uncertainty render it less useful when $R$ falls below 0.7. \par

\subsection{Computational Time}

The qualitative difference in the reliability of recovering the underlying CLR model between EM$_{is}$ and the traditional EM algorithm is stark. As shown in Section \ref{sec: algorithm performance}, the traditional EM algorithm is only able to recover the true CLR model in cases of extremely low complexity. Consequently, it is not particularly meaningful to compare the computational efficiency of the two methods. For this reason, only the computational complexity of EM$_{is}$ is presented in this section.

\begin{figure}[h]  
    \centering
    \includegraphics[width=1.0\textwidth]{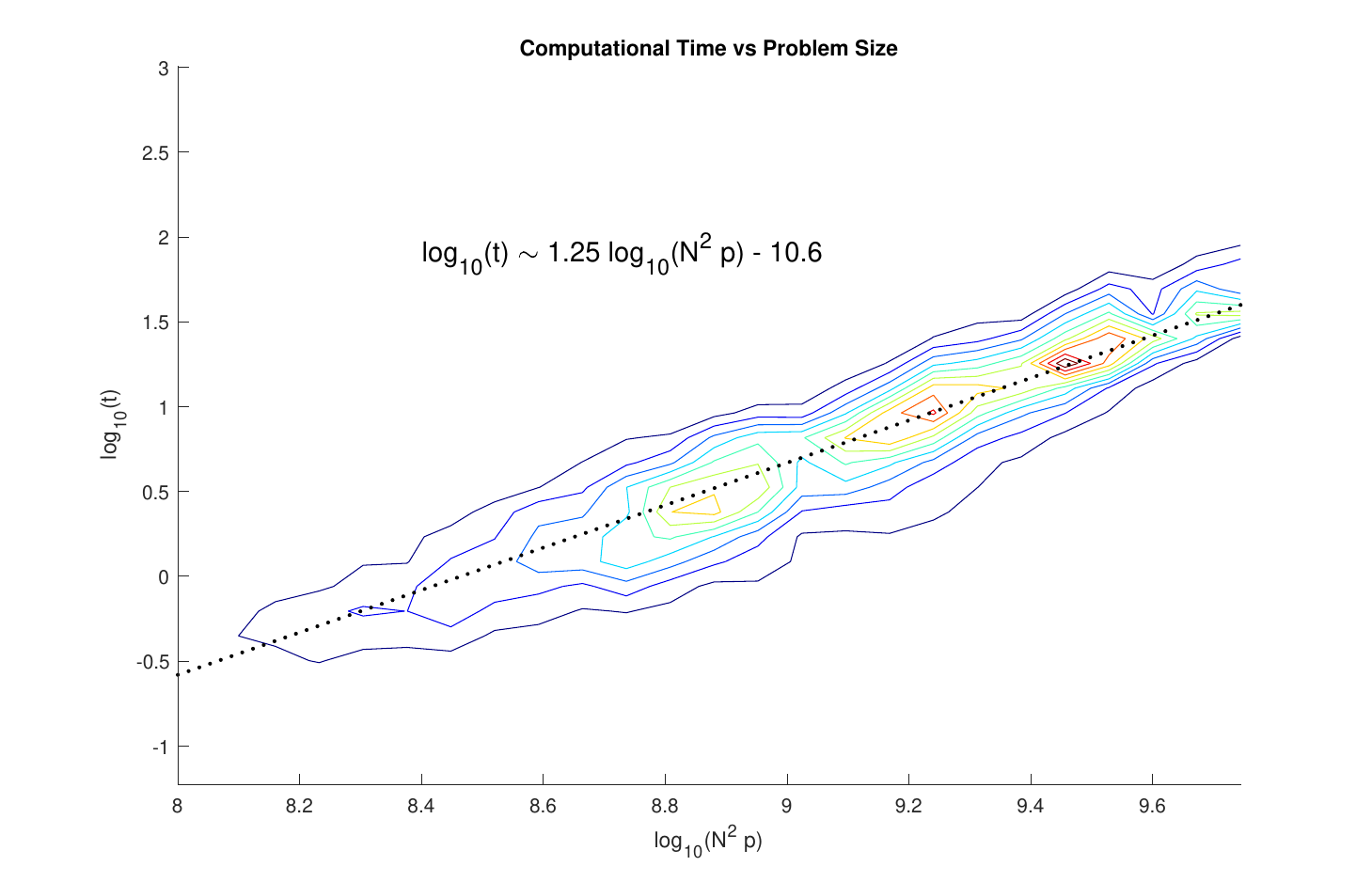}
    \caption{The contour plot of evaluation time density as a function of $N$ and $p$ of over 14000 independent problems. The empirical trend line shows that $\mathcal{O}(N^{2.50}p^{1.25}))$.} 
    \label{fig: computational time}
\end{figure}

The most computationally intensive step in the EM$_{is}$ algorithm is the estimation of $\hat{\hat{\beta_k}}$. A commonly reported complexity for this estimation step is $\mathcal{O}(K N_k p^2) \sim \mathcal{O}(N p^2)$. However, additional computational demands arise from subroutines within the Edge-Point K-flat and Center-Point Splitting algorithms. In particular, the estimation of the new $\beta_k$ in these methods involves operations such as principal component projection and eigen-decomposition, which are typically cited as having a similar complexity of $\mathcal{O}(K N_k p^2) \sim \mathcal{O}(N p^2)$. The nearest-neighbor search has the complexity on the order of $\mathcal{O}(N \log N)$. These estimates, however, reflect only the cost of a single iteration and do not account for the number of iterations required for convergence. Since the total iteration count depends on various factors such as noise level, cluster resolvability, data size, and dimensionality, this paper adopts an empirical approach to estimating the total computational time. These problems were all solved on on Matlab with parallel threads disabled, running on a Dell Precision 3640 Tower computer with Intel(R) Core(TM) i7-10700 CPU @ 2.90GHz and 32GB RAM. \par

Specifically, the total computation time $t$ for the 14000 independent problems used in Section \ref{sec: resolvability result} is modeled as a function of $N$ and $p$ using the equation $\log(t) = a_N \log(N) + a_p \log(p) + a_0$. It was found that $a_N = 2.50 \pm 0.03$, $a_p = 1.26 \pm 0.02$, and $a_0 = 10.58 \pm 0.13$ at the 95\% confidence interval. This can be simplified to $t \sim \mathcal{O}(N^{2.50}p^{1.25})$, shown in Figure \ref{fig: computational time}. Figure \ref{fig: computational time} plotted the contours of the density created by computational time with respect to the problem characteristics $N^2 p$. The stronger dependence on $N$ in the empirical result compared to the theoretical analysis suggests that the number of required iterations increases with the total sample size. The weaker dependence on $p$ relative to the predicted complexity of the regression and eigen-decomposition steps shows that the computational cost is dominated by the total iteration counts. Improvements in the computational efficiency should be sought by reduction in iteration counts. 

\subsection{Unbalanced Data and Corrupted Data}

\begin{figure}[h]  
    \centering
    \includegraphics[width=1.0\textwidth]{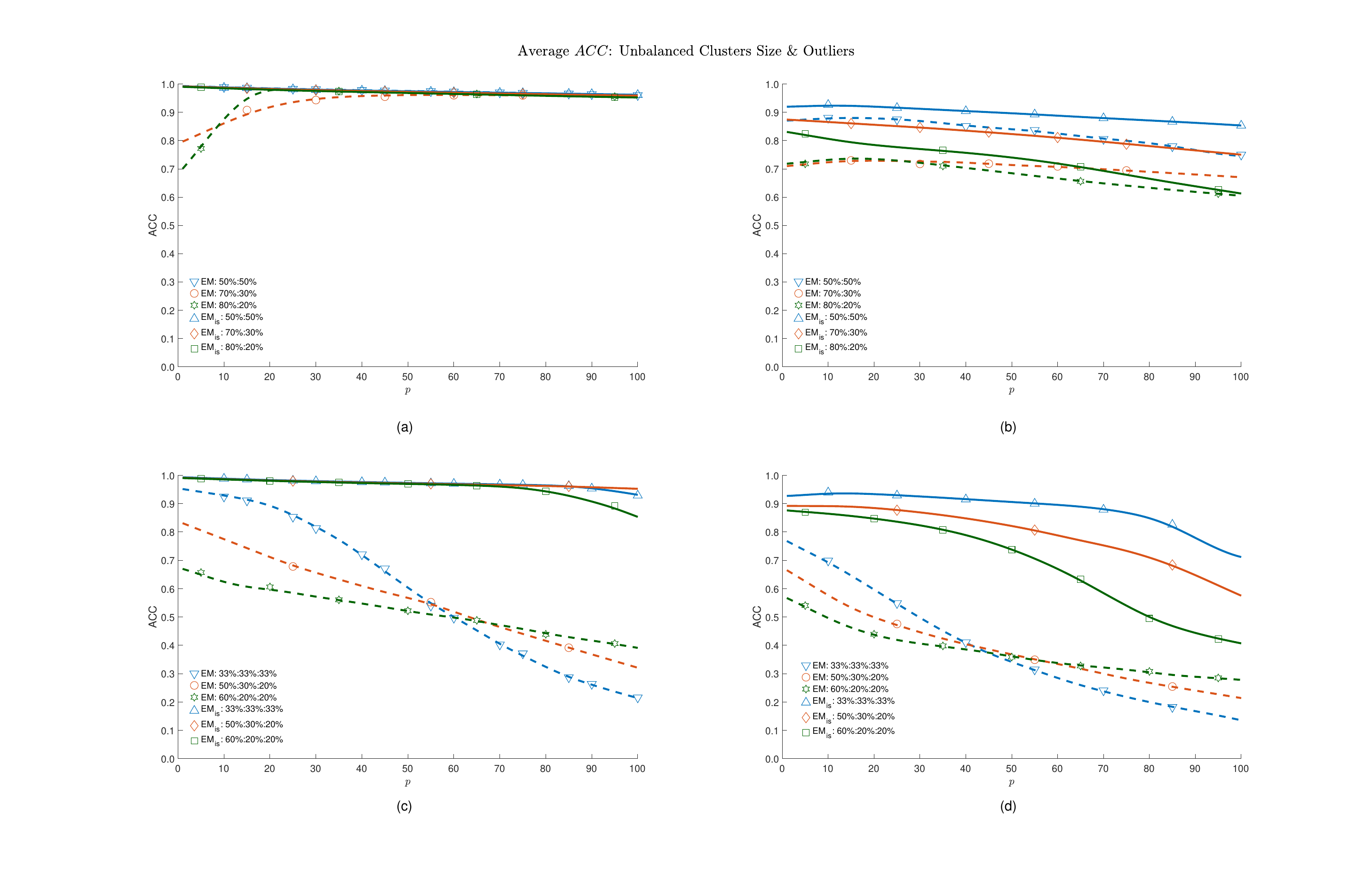}
    \caption{$ACC$ performance comparison between EM and EM$_{is}$ under cluster imbalance and the presence of outliers. (a) $K = 2$. (b) $K=2, 10\%$ outliers (c) $K = 3$. (d) $K=2, 10\%$ outliers.} 
    \label{fig: unbalanceoutlier}
\end{figure}

Unbalanced cluster sizes and the presence of corrupted data are expected to negatively affect the reliability of recovering the underlying CLR models. This section presents the performance of EM$_{is}$ relative to the traditional EM algorithm under such conditions. As discussed earlier, the effectiveness of EM$_{is}$, particularly the steps involved in proposing new estimates of $\hat{\hat{\beta}}$, depends on having a sufficient number of data points. In unbalanced datasets, the reliability of the regression analysis for the smallest cluster often becomes the weakest link, leading to a reduction in $ACC$ compared to balanced datasets. Figure \ref{fig: unbalanceoutlier} shows the average $ACC$ performance comparison between EM$_{is}$ and EM for scenarios with $K = {2, 3}$, $5 \leq p \leq 100$, $dp = 0.2$, $\eta = 0.1$, 10 restarts and a total sample size of 1000. The reported $ACC$ values are averaged over 1,000 independently generated random problems from the same class. To avoid cluttering the figure, only a subset of the data points is shown with markers. \par

The balanced datasets are represented by the $50\%:50\%$ curve in Figure \ref{fig: unbalanceoutlier}a and the $33\%:33\%:33\%$ curve in Figure \ref{fig: unbalanceoutlier}c. Both graphs show that the traditional EM algorithm is significantly affected by cluster imbalance, whereas EM$_{is}$ demonstrates greater robustness under such conditions. \par

Figure \ref{fig: unbalanceoutlier}b and Figure \ref{fig: unbalanceoutlier}d show the similar comparisons when the cluster size is not only unbalance, 10\% of the data are completely corrupted. Instead of generating $y$ value from cluster-specific generative model $X\beta_k + \varepsilon_k$ from the knowledge of $X$, the $y$ value of the corrupted data points are generated by a random number generator. The random number generator mimics the distributions of the uncorrupted data points to make sure that the corrupted data points are statistically indistinguishable from uncorrupted data point except the unexplained deviation from the generative model $X\beta_k$. The results show that the presence of corrupted data significantly affect our ability to recover the underlying CLR model but EM$_{is}$ is still very significantly more robust than the traditional EM algorithm. Comparing Figure \ref{fig: unbalanceoutlier}c and Figure \ref{fig: unbalanceoutlier}d also show the large impact corrupted data has on our ability to recover the underlying CLR model.In real-life data analysis scenario, this adds another dimension of consideration when the the algorithm fails to find a good CLR model.

\section{Conclusion}

This paper presents EM$_{is}$, an enhanced version of the standard Expectation-Maximization (EM) algorithm tailored for clusterwise linear regression (CLR). The authors propose several key improvements in EM$_{is}$, which make it more robust in handling challenges such as higher input dimensionality, increased cluster numbers, noise, and poor model resolvability. These advancements allow EM$_{is}$ to outperform traditional EM algorithms by increasing the probability of finding accurate solutions across a broader range of CLR problems.

\paragraph{Key Contributions}
\begin{enumerate}
\item \textbf{Improved Solution Quality}: EM$_{is}$ enhances the ability to find correct solutions under wide variety of problem characteristics, particularly in complex, high-dimensional, and noisy data environments. This improvement is qualitative, enabling EM$_{is}$ to solve classes of CLR problems previously unsolvable with traditional EM algorithms.
\item \textbf{Introduction of Resolvability and X-Predictability}: The paper introduces two new concepts.
\begin{itemize}
    \item \textbf{Resolvability}: Measures the quality of a CLR model without needing to know the true solution. Low resolvability may indicate potential overfitting or the given problem is not amenable to CLR modeling.    
    \item \textbf{X-Predictability} (XP): In the discussion of X-predictability, distinction has been made between CLR model that can be used to make prediction on new data and those that cannot.  A quantitative indicator of XP-ness has been proposed to helps gauge the trust-worthiness of the  CLR predictions and enables a more robust prediction reporting framework.
\end{itemize}
\item \textbf{Sample size requirement}: CLR modeling of high-dimensional data is data-hungry if the clusters overlap significantly. This requirement, overlooked in the literature, should be taken into consideration when CLR modeling is attempted. 
\end{enumerate}

These contributions are especially relevant for modern data analysis tasks, where high-dimensional data is common, and the need for reliable clustering and prediction models is critical. The improvements made in EM$_{is}$ empower practitioners to apply CLR analysis to a wider range of data types and increase the reliability of their results. Besides, EM$_{is}$ can be extended to more complex models, such as CWM and K-plane regression or other types of mixture models, replacing the standard EM framework in those contexts. However, further research is needed to address how EM$_{is}$ should be optimized to suit different problem formulations for best performance.

\newpage
\appendix
\setcounter{equation}{0}
\setcounter{figure}{0}

\section*{Appendix A: Resolvability Index}    \label{sec: appendix a}
    \renewcommand{\theequation}{A\arabic{equation}}

\subsection*{Derivation of $Z$}

\paragraph{Special case I}: Identical clusters. \\
Consider the situation of maximal overlap between clusters such that $\beta = \beta_1=\beta_2= \cdots =\beta_K$ and $\sigma = \sigma_1=\sigma_2= \cdots =\sigma_K$. This is the worst case scenario that we would like to let $Q$=1. Hence,  

\begin{eqnarray}
    1 =         Q &=& \int_{X} G(X) f(X) dX     \nonumber \\
    \Rightarrow Z &=& \int_{X} \int_{-\infty}^\infty \frac{e^{-\frac{1}{2}}(\frac{y-\tilde{X}\beta}{\sigma})^2}{(2\pi)^{K/2}\sigma^K} dy f(X) dX            \nonumber \\
                  &=& \frac{1}{\sqrt{K}(2\pi \sigma^2)^{\frac{K-1}{2}}}        \nonumber
\end{eqnarray}

In the previous expression, $\sigma$ was left undefined. This lack of specification is problematic, particularly given that $Z$'s dependence on $\sigma$ is undesirable. Furthermore, the above procedures do not provide a definition for $\sigma$ when $\sigma_k$ assumes different values. To address this, we impose an additional constraint on the equation by defining $\sigma$ as the geometric mean of $\sigma_k$. This definition allows for the calculation of $\sigma$ that is consistent with the conditions outlined in Special Case I.

\begin{eqnarray}
    \sigma  &=& \prod_{k=1}^K \sigma_k^{1/K}       \nonumber \\
    \Rightarrow Z^{-1} &=& (2\pi)^{\frac{K-2}{2}} \sqrt{K} \prod_{k=1}^{K} \sigma_k^{\frac{K-1}{K}}        \nonumber
\end{eqnarray}

\subsection*{Other special cases of interest}
A few other special (extreme) cases are presented below as a sanity check on the proposed $R$ (or $Q$) as a metric of CLR model quality. \\ 

\begin{figure}[h]
    \centering
    \includegraphics[width=1.0\textwidth]{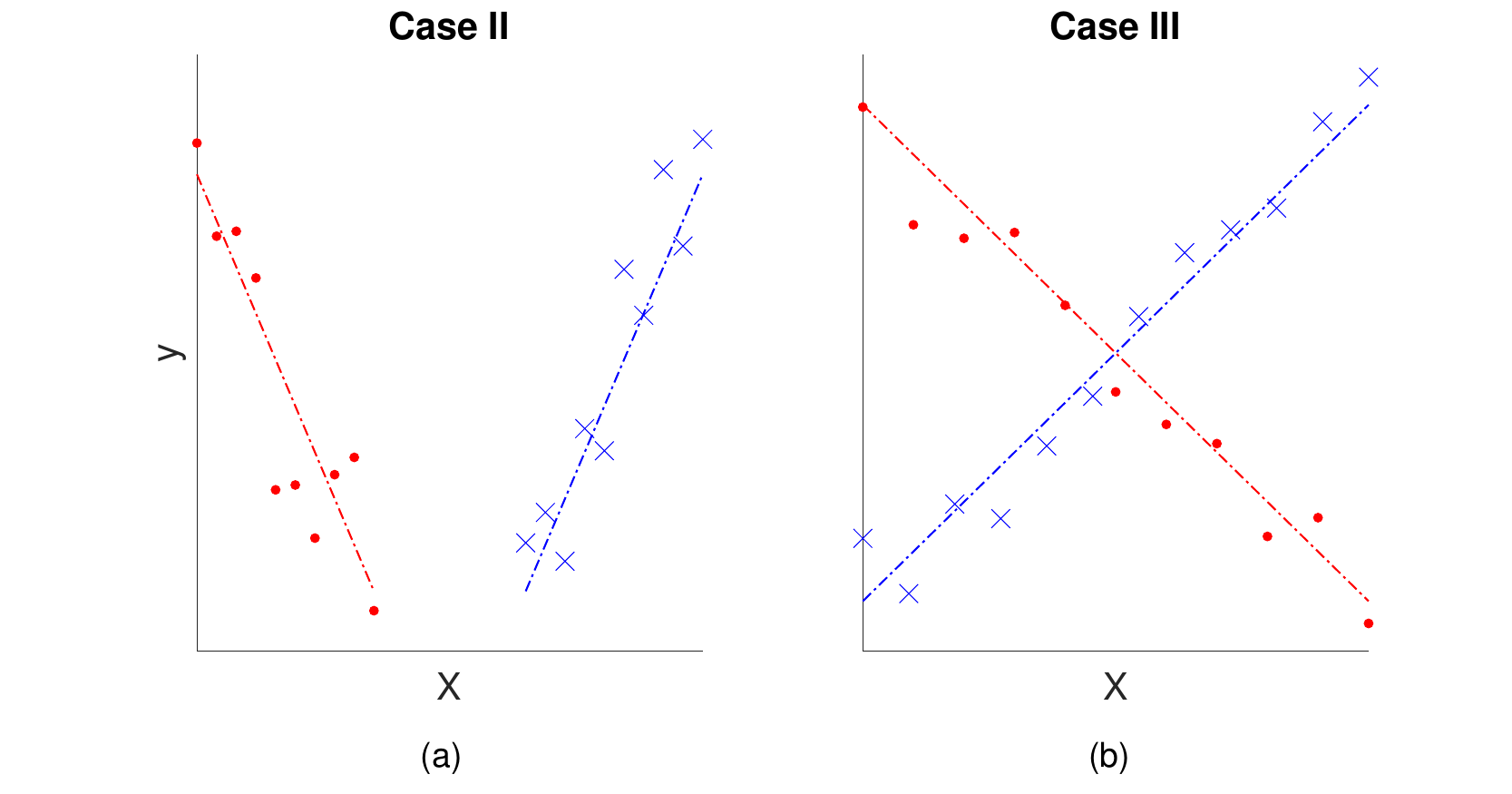}
    \caption{Illustration using one-dimensional examples. (a) Special case II, negligible overlap (b) Special case III, full overlap with different regression vectors.}  
    \label{fig: AppA01:case II and III}
\end{figure}

\paragraph{Special case II}: Negligible overlap between clusters.\newline
Negligible overlap guarantees that $G(X) \approx 0$ at every point $X$. This leads to $Q \approx 0 \Rightarrow R \approx 1$. This is an extreme case of perfectly XP data elaborated in Section \ref{sec: making predictions}. The analysis of special case II shows that $R$ is very useful but also indicated a very subtle weakness of $R$, i.e. there is a special class of overfitting that $R$ is unable to detect. For example, set $K=2$ for a dataset with only one cluster and $\beta_1=\beta_2$ and $\sigma_1=\sigma_2$ but the algorithm partition $X$ into two disjoint sets. We will still find $R \approx$1 even though the two clusters are actually describing the same cluster. Fortunately, empirical testing of the EM algorithms on CLR problems shows that this situation is not a cause of worry because it is highly unstable and almost impossible to attain in real life. Slight perturbation in data will cause one of the clusters to gain upper-hand in attracting data points and eventually, through iteration in EM loop, renders the weaker cluster an empty set. \newline

\paragraph{Special case III}: Overlapped domain $X$ with $\sigma = \sigma_1=\sigma_2= \cdots =\sigma_K$ but $\beta_i \neq \beta_j$. \newline
This is an extreme case of perfectly non-XP data. Some MLR problem formulations and algorithms implicitly assumed the data is of this form. Assuming all cluster having the same uncertainty $\sigma$ simplifies the derivation without compromising the main conclusion. 

\begin{eqnarray}
    Q &=& \sqrt{\frac{K}{2\pi \sigma^2}} \int_{X} \int_{-\infty}^\infty e^{-\frac{1}{2 \sigma^2}\sum_{k=1}^{K} (y-\tilde{X}\beta_k})^2 dy f(X) dX     \nonumber \\
      &=& \sqrt{\frac{K}{2\pi \sigma^2}} 
      \int_{X} \int_{-\infty}^\infty e^{-\frac{K}{2 \sigma^2} (y-\frac{1}{K} \sum_{k==1}^K \tilde{X}\beta_k )^2} e^{\frac{-\alpha}{2 \sigma^2}} dy f(X) dX     \nonumber 
\end{eqnarray}

where $\alpha = \sum_{k=1}^K(\tilde{X}\beta_k)^2 - \frac{1}{K}(\sum_{k=1}^K \tilde{X}\beta_k)^2 \ge 0$ from Cauchy-Schwarz inequality $(\sum_{k=1}^K(\tilde{X}\beta_k)^2)(\sum_{k=1}^K 1^2) \ge (\sum_{k=1}^K(\tilde{X}\beta_k))^2$. Continue the simplification yields:
\begin{eqnarray}
    Q &=& \int_{X} e^{-\frac{\alpha}{2 \sigma^2}} f(X) dX   \nonumber
\end{eqnarray}

The last integral leads to conclusion $Q$ as a function of the ratio $\frac{\alpha}{\sigma^2}$ and 0 $\leq$ Q $\leq$ 1. $\alpha$ is determined by the angular separation between the clusters regression vectors $\beta_k$. From the definition, $\alpha$ gets smaller when $\beta_k$ are more similar to each other. On the other hand, $\sigma$ measures the data acquisition noise or inherent uncertainty of the regression model. The ratio $\frac{\alpha}{\sigma^2}$ measures the relative angular separation between the clusters against the level of noise. The model has low resolvability when either $\alpha \rightarrow 0$ or $\sigma \rightarrow \infty$.In other words, the cluster separation relative to the noise level is the principal determining the resolvability. This observation agrees with the expected behavior of the $R$ as a quality metric of a CLR model. \newline

\paragraph{Special case IV}: Fully overlapped domain $X$ with $\beta = \beta_1=\beta_2= \cdots =\beta_K$ but $\sigma_i \neq \sigma_j$. \\
For simplicity, we limit the discussion to $K$=2, leading to $\beta = \beta_1=\beta_2$ and $\sigma_1 \neq \sigma_2$. Without lost of generality, let's further assume $\sigma_1 \ge \sigma_2$. $Q$ in this case is given by the following expression:

\begin{eqnarray}
    Q &=& \sqrt{\frac{2}{2\pi \sigma_1\sigma_2}} \int_{X} \int_{-\infty}^\infty e^{-\frac{\sigma_1^2+\sigma_2^2}{2 \sigma_1^2 \sigma_2^2} (y-\tilde{X}\beta)^2 } dy f(X) dX     \nonumber \\
    &=& \sqrt{\frac{2\sigma_2/\sigma_1}{1+\sigma_2^2/\sigma_1^2}} \nonumber    
\end{eqnarray}

We have 0 $\leq$ Q $\leq$ 1; $\lim_{\sigma_2/\sigma_1 \to 0}R = 1$ while $\lim_{\sigma_2/\sigma_1 \to 1}R = 0$. Just like the note on special case II, this is not a stable configuration unless the data exhibit long-tail distribution where mixture of Gaussian distribution offers a better approximation. The decision weather to accept the resultant model will rest on the hand of the user. 

\begin{figure}[h]
    \centering    
    \includegraphics[width=0.7\textwidth]{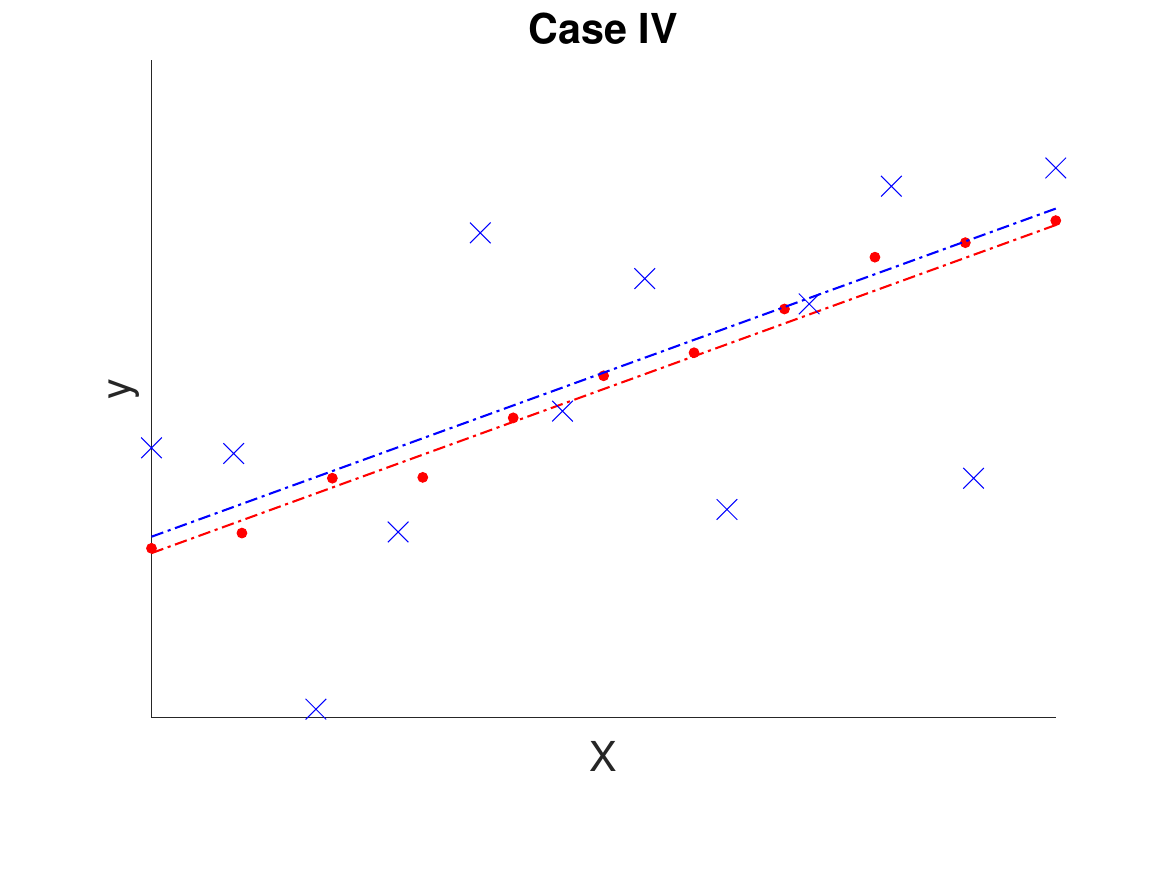}
    \caption{Illustration using one-dimensional examples. Special case IV, full overlap with identical regression vectors but different level of uncertainty.}  
    \label{fig: AppA02:case IV}
\end{figure}

\section*{Appendix B: Coordinate transformation}    \label{sec: appendix b}

\subsection*{Forward coordinate transformation $\mathscr{T}$}  

\begin{algorithm}[h!]    \label{alg: T}
    \caption{Forward coordinate transformation $\mathscr{T}$} 
    \KwIn{$X_s \in \mathbb{R}^{M \times p }$, $y_s \in \mathbb{R}^{M \times 1}$, $\beta_0 \in \mathbb{R}^{(p+1) \times 1}$}
    \KwOut{$P \in \mathbb{R}^{M \times (D-1)}$, $\alpha_0 \in \mathbb{R}^{D \times 1}$}
    
    \Parameters{$\theta_{PCA}$}
    
    \Definitions{ $Z = [X_s,y_s]$;  $\mu_Z = [\mu_X,\mu_y]$; $\sigma_Z = [\sigma_X,\sigma_y]$; $\Sigma = diag(\sigma_Z)$; $\Lambda = diag(\lambda_1,...,\lambda_{p+1})$ }
    
    Centering and whitening: $ \hat{Z} = (Z-\mu_Z)\Sigma^{-1}$ \\
    Eigen-decomposition: $\frac{1}{M} (\hat{Z}^{T} \hat{Z}) = \tilde{V} \Lambda \tilde{V}^{T}$  where $\lambda_1 \geq ... \geq \lambda_{p+1}$ \\
    Partition $\tilde{V}=[\overset{D-1}{\overleftrightarrow{V}},V_{0}]$ such that $\frac{\lambda_{i \geq D}}{\lambda_1} < \theta_{PCA}$, retain only $V$ \\
    
    Principle component projection: $P = \hat{Z}V $\\
    $\check{\alpha}_0 = V^T \Sigma [\check{\beta_0},-1]^T$ \\
    $\beta_0$ in projected space:  $\alpha_0 = [0, \check{\alpha}_0]^T$ 
\end{algorithm}

Step 4.1 whitening will remove the effect of natural scale differences between different predictors on the subsequent metric-based operations. Step 4.2-4.4 are standard principal component projection. It de-correlates the components of $P$ while removing small noisy components through $\theta_{PCA}$. A small value can be chosen for $\theta_{PCA}$ but the exact value is not critical. A default value of $\theta_{PCA}=10^{-8}$ will be used through out this paper. Step 4.5 and 4.6 calculates the representation of $\beta_0$ in the new coordinate system. The derivation is shown below: \par

Consider a hyperplane define by $\beta_0$:
\begin{eqnarray}
            y   &=& \tilde{X}\beta_0 = \beta_{00} + X \check{\beta} \nonumber \\
  0 &=& \beta_{00} + Zb \quad\quad \text{where} \quad b = \begin{bmatrix} \check{\beta} \\ -1 \end{bmatrix} \nonumber \\
    &=& \beta_{00} + (\mu_Z + \hat{Z} \Sigma)b \nonumber \\
    &=& (\beta_{00}+\mu_Z b) + \hat{Z} \Sigma b \nonumber \\
    &=& (\beta_{00}+\mu_Z b) + \hat{Z} \tilde{V} \tilde{V}^T \Sigma b  \nonumber \\
    &=& (\beta_{00}+\mu_Z b) + \hat{Z} (V V^T + V_0 V_0^T) \Sigma b  \nonumber
\end{eqnarray}

Note that $\beta_{00}+\mu_Z b = 0$ is the restatement of the orignal hyperplane equation at the centroid. Define $[P,P_0] = \hat{Z}[V,V_0]$. $\vert P_0 \vert \ll \vert P \vert$ by definition if $\theta_{PCA}$ is chosen to be a very small value. Substitute $P$ and $P_0$ into the equation yields:
\begin{eqnarray}
  0 &=& P V^T \Sigma b + P_0 V_0^T \Sigma b  \approx P V^T \Sigma b \nonumber
\end{eqnarray}
Discard $P_0 V_0^T \Sigma b $ term because its contribution is negligible. Then the equation of the equivalent hyperplane in the projected space is given by 
\begin{eqnarray}
  0 &=& P (V^T \Sigma b) = P \alpha_0   \nonumber \\
  \alpha_0  &=& V^T \Sigma b            \nonumber  \\
  &=&  V^T \Sigma \begin{bmatrix} \check{\beta} \\ -1 \end{bmatrix}  \nonumber     
\end{eqnarray}

\subsection*{Inverse coordinate transformation $\mathscr{S}$}

\begin{algorithm}[h]    \label{alg: S}
\caption{Inverse coordinate transformation $\mathscr{S}$} 
\KwIn{$\alpha = [\alpha_1,\alpha_2]$}
\KwOut{$\beta_{new}$}
\Parameters{from $\mathscr{T}(\cdot)$: $\mu_Z$, $\Sigma$, $V$}
    \For{$k \gets 1$ \KwTo $2$}{
    $ b_k = \Sigma^{-1} V \check{\alpha}_k $ \\
    Partition $ b_k = [\overset{p}{\overleftrightarrow{b_{k;X}}}, \overset{1}{\overleftrightarrow{b_{k;y}}}]^T$ \\
    $ \beta_{k0} = (\mu_Z b_k-\alpha_{k0})/b_{k;y} $ \\
    $\check{\beta}_{k} = -b_{k;X} / b_{k;y}$ \\
    }
    $\beta_{new} = [\beta_1,\beta_2]$
\end{algorithm}

Equations in $\mathscr{S}$ are derived below:
\begin{eqnarray}
    0 &=& C_k \alpha_k              \nonumber \\
      &=& \begin{bmatrix} 1 && \hat{Z} V \end{bmatrix}\begin{bmatrix} \alpha_{k0} \\ \check{\alpha}_k \end{bmatrix}                  \nonumber \\
      &=& \alpha_{k0} + (Z-\mu_Z)\Sigma^{-1} V  \check{\alpha}_k         \nonumber \\
      &=& (\alpha_{k0} - \mu_Z b) + X b_{k;X} + y b_{k; y}              \nonumber \\
      \implies y &=& -\frac{\alpha_{k0}-\mu_Z b}{b_{k;y}} - \frac{X b_{k;X}}{b_{k;y}}           \nonumber \\
      \implies \tilde{X} \beta &=& \tilde{X} \quad \begin{bmatrix} \frac{-1}{b_{k;y}} (\alpha_{k0}-\mu_Z b) \\  \frac{-1}{b_{k;y}} b_{k;X} \end{bmatrix}      \nonumber
\end{eqnarray}

\section*{Appendix C: Elite Recombination}    \label{sec: appendix c}

Algorithm \ref{alg: EliteRecombination} shows the elite recombination procedures. The hyperparameter values given in parenthesis in Algorithm \ref{alg: EliteRecombination} are value used to generate the results in this paper. The Elite, $\mathscr{E}$, stores best solution sets over the EM iteration. The solution set consists of the information required to re-enact the CLR model including $\hat{\beta}$, $\hat{\sigma}$, $w$ and miscellaneous statistics such as various metrics of merits, estimated prior of each cluster, resolvability and time/iterations to convergence. The length of Elite, $|\mathscr{E}|$, is the number of best solutions stored. Larger $|\mathscr{E}|$ makes the recombination method more powerful in large search space scenario such as large $K$ and large $p$. However,  larger $|\mathscr{E}|$ also incur higher computational cost. The results in this paper are all generated using $|\mathscr{E}|=5$. \par

\begin{algorithm}[h]    \label{alg: EliteRecombination}
\caption{Elite recombination} 
\KwIn{$\mathscr{E}$}
\KwOut{$\beta_{new}$}
\Parameters{$T_{s_1}(0.5)$, $T_{s_2}(0.8)$, $T_{s_3}(\frac{1}{3K})$,$\textnormal{MaxLen}(7)$}

For every solution pair with $w$ correlation $> T_{s_1}$, remove the weaker solution from $\mathscr{E}$ \\
\eIf{$|\mathscr{E}|$ = 1}{
    Remove the smallest cluster from the solution \\
    Use procedures in Section \ref{sec: new beta} to generate $\beta_{new}$ 
    }{
    Break $\mathscr{E}$ into collection of cluster proposals $\mathscr{C}$ \\
    From $\mathscr{C}$: \\
    \Indp
    Remove weaker clusters with correlation $> T_{s_2}$ \\
    Remove small clusters size $< T_{s_3}$  \\
    \Indm
    \eIf{$|\mathscr{C}| < K$}{
        Combine remaining clusters in $\mathscr{C}$ \\
        Use procedures in Section \ref{sec: new beta} to generate $\beta_{new}$
    }{
        $L = min(|\mathscr{C}|,\textnormal{MaxLen})$
        Select  $\binom{L}{K}$ random combinations of cluster in $\mathscr{C}$ \\
        \ForEach{\textnormal{cluster combination}}{
        Re-estimate $\hat{\beta}$ \\
        Evaluate regression error
        }
        Return the combination with the smallest regression error
    }
}
\end{algorithm}

Step 7.1 removes strongly correlated elite solutions to reduce the number of combinations in subsequent steps. Two solutions are considered highly correlated if the correlation between their cluster assignments $w$ exceeds the threshold $T_{s_1}$. The weaker solution is defined as the CLR model with the larger regression error. In Steps 7.2–7.4, if only one solution remains in $\mathscr{E}$ after filtering indicating a lack of diversity among the best solutions, a new proposal is generated by perturbing the remaining solution. On the other hand, if multiple sufficiently distinct solutions remain, they are decomposed into their constituent clusters $\mathscr{C}$ in Step 7.6 to enable new cluster proposal recombination. Again, to reduce the computation load, the clusters in $\mathscr{C}$ are removed through heuristics such as relative cluster size and similarity measured through  inter-cluster correlation. The proposals remaining in $\mathscr{C}$ are combined to form $K$-cluster solutions in Steps 7.10-7.20. The combinations that achieve least regression error will be returned to re-initialize the EM iterations. \par

The above descriptions show that Algorithm \ref{alg: EliteRecombination} filters through many potential cluster combinations to retain only one most promising candidate to re-initialize the main EM iterations. Increasing the number of proposals offers only marginal gains in the likelihood of discovering promising combinations, while the computational cost grows exponentially. In the absence of a reliable predictor to evaluate the quality of a cluster proposal—a problem that is NP-hard—the use of heuristics in Algorithm \ref{alg: EliteRecombination} always carries the risk of prematurely discarding a good proposal. However, it is even more likely that, at the time of recombination, some hidden clusters have not yet been discovered by EM${is}$ and are therefore not present in $\mathscr{E}$. In such cases, expending excessive computational effort on evaluating combinations is counterproductive. EM${is}$ favors more frequent restarts while aggressively limiting the number of combinations considered in each re-initialization, thereby increasing the chance of uncovering all underlying cluster structures. The efficiency implications of these design choices are left for future research, allowing this paper to remain focused on the conceptual contributions of the EM$_{is}$ algorithm.

%% References
\bibliographystyle{elsarticle-harv} 
\bibliography{Reference}

\end{document}